\documentclass[preprint,pra,showpacs,showkeys,floatfix]{revtex4}
\usepackage{graphicx}
 \usepackage{amsmath}
 \usepackage{amsfonts}

\begin{document}

\title{WavePacket: A Matlab package for numerical quantum dynamics.\\
II: Open quantum systems, optimal control, and model reduction} 

\author{Burkhard Schmidt}
\email{burkhard.schmidt@fu-berlin.de}
\affiliation{
Institut f\"{u}r Mathematik, Freie Universit\"{a}t Berlin\\ Arnimallee 6, D-14195 Berlin, Germany}

\author{Carsten Hartmann}
\email{carsten.hartmann@b-tu.de}
\affiliation{
Institut f\"{u}r Mathematik, Brandenburgische Technische Universit\"{a}t\\Konrad-Wachsmann-Allee 1, D-03046 Cottbus, Germany}

\date{\today}

\begin{abstract}
WavePacket is an open-source program package for numeric simulations in quantum dynamics. 
It can solve time-independent or time-dependent linear Schr\"{o}dinger and Liouville-von Neumann-equations in one or more dimensions. 
Also coupled equations can be treated, which allows, e.g., to simulate molecular quantum dynamics beyond the Born-Oppenheimer approximation.
Optionally accounting for the interaction with external electric fields within the semi-classical dipole approximation, WavePacket can be used to simulate experiments involving tailored light pulses in photo-induced physics or chemistry.
Being highly versatile and offering visualization of quantum dynamics 'on the fly', WavePacket is well suited for teaching or research projects in atomic, molecular and optical physics as well as in physical or theoretical chemistry.
Building on the previous Part I [Comp. Phys. Comm. \textbf{213}, 223-234 (2017)] which dealt with closed quantum systems and discrete variable representations, the present Part II focuses on the dynamics of open quantum systems, with Lindblad operators modeling dissipation and dephasing.
This part also describes the WavePacket function for optimal control of quantum dynamics, building on rapid monotonically convergent iteration methods. 
Furthermore, two different approaches to dimension reduction implemented in WavePacket are documented here.
In the first one, a balancing transformation based on the concepts of controllability and observability Gramians is used to identify states that are neither well controllable nor well observable.
Those states are either truncated or averaged out. 
In the other approach, the H2-error for a given reduced dimensionality is minimized by H2 optimal model reduction techniques, utilizing a bilinear iterative rational Krylov algorithm.

The present work describes the MATLAB version of WavePacket 5.3.0 which is hosted and further developed at the Sourceforge platform, where also extensive Wiki-documentation as well as numerous worked-out demonstration examples with animated graphics can be found. 

\end{abstract}

\maketitle

\section*{Program Summary}

\begin{description}
\item [Program Title]~\\ \textsc{WavePacket}
\item [Licensing provisions]~\\ GPLv3
\item [Programming language]~\\ \textsc{Matlab}
\item [Journal reference of previous version]~\\ Comput. Phys. Comm. \textbf{213} (2017), 223.
\item [Does the new version supersede the previous version?]~\\ The previous article focused on the treatment of closed quantum systems by discrete variable representations and implementation of various numerical algorithms for solving Schr\"{o}dinger's equations.
Complementary to that, the present second part is concerned with open quantum systems and optimal control by external fields.
In addition, two approaches to dimension reduction useful in modeling of quantum control are described.
\item [Reasons for the new version]~\\ The reason for having a second article on the WavePacket software package lies in the fact that a complete description of the package would have exceeded the scope of a regular article. Several significant features of the \textsc{WavePacket} package are introduced here which could not be mentioned in the first article, due to length constraints.
\item [Summary of revisions]~\\
Here we describe the numerical treatment of open quantum systems dynamics modeled by Lindblad master equations. 
Moreover, we explain the \textsc{WavePacket} functions for optimal control simulations, both for closed and open quantum systems. 
To address the problem of computational effort, two strategies for model reduction have been included.
\item [Nature of problem]~\\
  Dynamics of closed and open systems are described by Schr\"{o}dinger or Liouville-von Neumann equations, respectively, where the latter ones will be restricted to the Lindblad master equation.
	Emphasis is on the interaction of quantum system with external electric fields, treated within the semi-classical dipole approximation.
	Quantum optimal control simulations are used for the design of tailored fields achieving specified targets in quantum dynamics. 
	With these features, \textsc{WavePacket} can be instrumental for the simulation, understanding, and prediction of modern experiments in atomic, molecular and optical physics involving temporally shaped fields.
\item [Solution method]~\\
   Representing state vectors or reduced density matrices in a discrete basis, Schr\"{o}dinger or Liouville-von Neumann equations are  cast into systems of ordinary differential equations. Those are treated by self-written or \textsc{Matlab}'s built-in solvers, the latter ones offering adaptive time-stepping.
	The optimal control equations are solved by the rapid monotonically convergent iteration methods developed by  Zhu, Rabitz, Ohtsuki and others.
	In order to reduce the dimensionality of large scale control problems, the balanced truncation method as well as H2-optimal model reduction approaches are available in \textsc{WavePacket}.
\item [Additional comments including restrictions and unusual features]~\\
  The \textsc{WavePacket} program package is rather easy and intuitive to use, providing visualization of quantum dynamics 'on the fly'. 
	It is mainly intended for low-dimensional systems, typically not exceeding three to five degrees of freedom.
	Detailed user guides and reference manuals are available through numerous Wiki pages hosted at the \textsc{SourceForge} platform where also a large number of well documented demonstration examples can be found.
\end{description}


\section{Introduction}
\label{sec:intro}

The evolution of ultrafast experimental techniques, mainly triggered by advances in generating short intense laser pulses in the late 20th century, has been a strong motivation for studying quantum mechanics also from the time-dependent point of view \cite{May:00a,Schleich:01a,Tannor:04a}.
Nowadays, experiments using tailored laser fields are often accompanied by quantum dynamical simulations resulting in substantial progress in the fields of atomic and molecular physics \cite{Schleich:01a,Grossmann2008}, femtochemistry \cite{Zewail2000,Tannor:04a} and even femtobiology \cite{Sundstrom2008}. 
Concepts developed in these fields are also instrumental in quantum information theory; for approaches to quantum computing in molecular physics see, e.~g. Refs.~\cite{Vivie:07a,Zhu2013,Kais2014}.
It can be expected that the control of quantum systems may lead to a variety of potential quantum technologies in the future \cite{Glaser2015}.

Despite of the obvious importance of quantum dynamical simulations, general-purpose and freely available simulation software is still rather scarce; notable exceptions being the MCTDH program package which specializes in weakly coupled, high-dimensional systems \cite{Beck2000}, or the nearly linearly scalable TDDVR package \cite{Khan2014}, both of which are mainly used in the context of quantum molecular dynamics.
Other software packages more commonly used in the physics community include \textsc{QuTiP} for the dynamics of open quantum systems \cite{Johansson2012, Johansson2013}, the \textsc{FermiFab} toolbox for many-particle quantum systems \cite{Mendl2011}, and the \textsc{QLib} platform for numeric optimal control \cite{Machnes2011}. 

In this work we continue the description of the \textsc{Matlab} version of our \textsc{WavePacket} software for numeric quantum dynamics which we decided to split into two articles, due to length constraints.
In Part I we have introduced this general software package, with regard to its use for closed quantum systems \cite{BSchmidt:75}, i.~e., mainly the solution of Schr\"{o}dinger equations (SE). 
These parts of \textsc{WavePacket} are based on describing wave functions and  operators in finite basis representations (FBRs) and/or associated discrete variable representations (DVRs) \cite{Light:85a,Light:00a}.
This FBR/DVR approach allows to cast the time-independent Schr\"odinger equation (TISE) into an eigenvalue problem which is solved by the \textsc{WavePacket} function \texttt{qm\_bound}.
In close analogy, the time-dependent Schr\"odinger equation (TDSE) is solved in a partial differential equation (PDE) setting by a variety of propagation methods implemented in the function \textsc{WavePacket} \texttt{qm\_propa}.
The efficiency of both approaches relies on  fast transformations between DVRs and FBRs, the most prominent example of which being fast Fourier transforms for use with plane wave FBRs \cite{Leforestier:91a}.
Finally, among the \textsc{WavePacket} functions introduced in Part I there is also a visualization tool \texttt{qm\_movie} which can be used to generate different types of animated graphics, as well as the auxiliary functions \texttt{qm\_setup} and \texttt{qm\_cleanup} to initialize and finalize simulation protocols.

Another emphasis of Part I is on the manipulation of quantum systems by external electric fields.
The interactions are treated within the framework of the semi-classical dipole approximation, hence making \textsc{WavePacket} especially suitable for simulating experiments in photophysics or photochemistry where shaped field pulses are used to alter, and ultimately to control, the dynamics of quantum systems.
Yet another feature of \textsc{WavePacket} is that it can treat coupled (multi-channel) SEs occuring, e.~g., for systems with slow and heavy degrees of freedom, such as nuclei and electrons in molecular systems.
Using \textsc{WavePacket}, the quantum dynamics of such systems can be treated beyond the Born-Oppenheimer (adiabatic) approximation, including situations where the dynamics is typically dominated by non-adiabatic transitions occurring near (avoided) crossings or conical intersections of potential energy curves or surfaces, respectively \cite{Baer:06a,Domcke:04a}.

\begin{figure}
	\caption{Flow chart of the \textsc{WavePacket} functions, also indicating the names of the files used for exchange of data between them. The left dashed red line indicates the border between the descriptions of quantum dynamics by partial differential equations (PDEs, using discrete variable representation) and by ordinary differential equations (ODEs, using energy representation). The right dashed red line marks the separation between quantum mechanics (QM) and dimension reduction and optimal control theory (OCT).}
	\includegraphics[width=10cm]{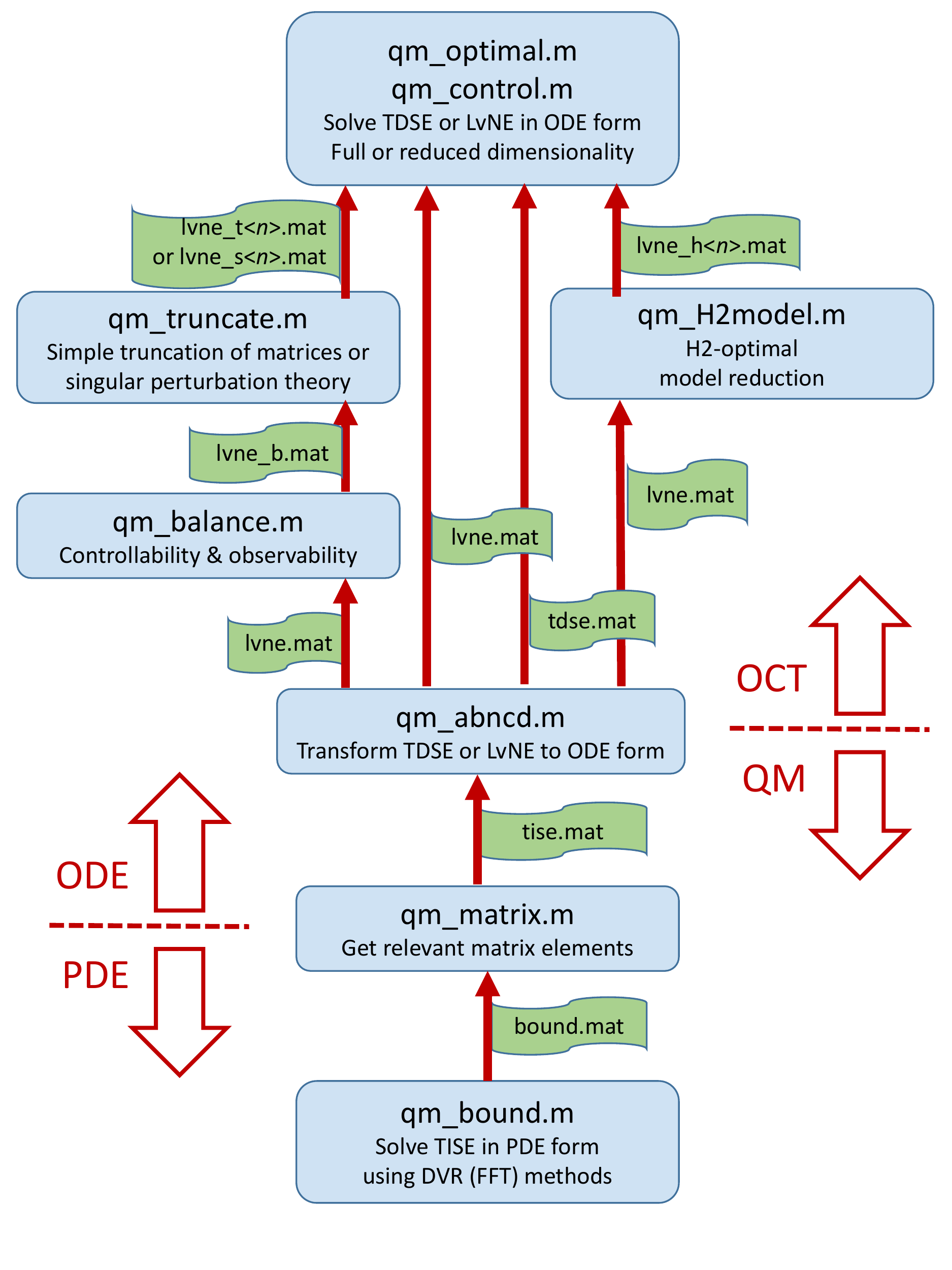}
	\label{fig:flowchart}
\end{figure}

The present, complementary Part II extends the previous work of Ref.~\cite{BSchmidt:75} by describing the use of \textsc{WavePacket} for simulations of open quantum systems modeled in terms of Liouville-von Neumann equations (LvNE) \cite{Weiss:99a,Breuer:02a}. 
Here, we will restrict ourselves to Lindblad-Kossakowski models for dissipation and dephasing of quantum systems interacting with a thermal bath \cite{Kossakowski1972,Lindblad1976}.
In order to treat closed and open quantum systems on an equal footing, it is advantageous to change from DVR ("coordinate") and/or FBR ("momentum") representations to an eigen ("energy") representation.
Then the corresponding equations of closed (TDSE) and open (LvNE) systems become sets of coupled ordinary differential equations (ODEs).
In fact, all \textsc{WavePacket} functions introduced in the present Part II are based on an energy representation in terms of eigenenergies and eigenfunctions which can be obtained, e.~g. using \texttt{qm\_init} and \texttt{qm\_bound}, see Secs.~\ref{sec:qm_init} and \ref{sec:qm_bound}.
The further course of a typical workflow is shown in the flow chart in Fig.~\ref{fig:flowchart}.
Within \textsc{WavePacket}, the actual change of representation is carried out in our function \texttt{qm\_matrix}, see Sec.~\ref{sec:qm_matrix}.
This is followed by function \texttt{qm\_abncd} which sets up the ODE formulations of the TDSE or the LvNE.
In doing so, we make use of the fact that quantum dynamics of driven systems within the semi-classical dipole approximation displays many formal analogies with control theory of bilinear systems.
Note that the name \texttt{qm\_abncd} alludes to the usual convention of denoting the system matrices as $A,B,N,C,D$, see Sec.~\ref{sec:qm_abncd}.

Once the system matrices have been set up, the TDSEs or the LvNEs are solved numerically by using conventional ODE solvers, which is realized in our code \texttt{qm\_control} described in Secs.~\ref{sec:qm_control}.
Moreover, since version 5.2.0 released in 2016, also optimal control techniques have been implemented in \textsc{WavePacket} where the time-dependence of optimal control fields is determined automatically, striving at optimizing certain user-defined control targets \cite{Judson:92a,Rabitz2000,Rabitz:03a}, often subject to constraints arising due to laboratory technologies. 
Typical examples for such constraints are limitations in the intensity of available light pulses.
With our function \texttt{qm\_optimal} described in Sec.~\ref{sec:qm_optimal}, we present a code for optimization of various types of targets, subject to field constraints, which builds on the rapidly converging methods introduced in Refs.~\cite{Zhu1998,Zhu1998a,Ohtsuki1999,maday2003,Ohtsuki2004,Werschnik2007}. 
Due to the general formulation in terms of the system matrices used in control theory, the \textsc{WavePacket} function \texttt{qm\_optimal} can be used both for closed and open quantum systems, or even for control problems from completely different sources, e.~g. classical Langevin dynamics \cite{Zhang2014} or Fokker-Planck dynamics \cite{BSchmidt:77}.

The main obstacle when simulating the control of multi-dimensional quantum systems is the exponential growth of the number of quantum states with the number of the degrees of freedom. 
Even for relatively small systems, this can lead to an exceedingly high computational effort, especially for open quantum systems (LvNE) dynamics where the size of the system matrices scales quadratically with the number of quantum states involved, which makes dimension reduction highly desirable. 
For such cases, \textsc{WavePacket} offers a balancing transformation reconciling the concepts of controllability and observability \cite{BSchmidt:61}.
As described in Sec.~\ref{sec:qm_balance}, the function \texttt{qm\_balance} aims at constructing states which are both controllable and observable.
The remaining states, i.~e. those which are neither well controllable nor well observable, do not contribute notably to the input-output behavior of a control system. 
Hence, the WavePacket function \texttt{qm\_truncate} can be used to eliminate those states, see Sec.~\ref{sec:qm_truncate}.
This is either achieved by a simple truncation or, in analogy with systems displaying fast and slow degrees of freedom, by averaging them out, based on singular perturbation theory \cite{hartmann2010c,hartmann2010e,BSchmidt:77}.
An alternative approach to model reduction will be given in Sec.~\ref{sec:qm_H2model}.
The function \texttt{qm\_reduce} serves to minimize the residual ${\mathcal H}_2$-error quantifying the deviation between the full system and a system of (given) reduced dimensionality, utilizing a bilinear iterative rational Krylov algorithm \cite{Breiten2010,Benner2012,BSchmidt:77}.

\section{qm\_init}
\label{sec:qm_init}

\subsection{Initialize WavePacket simulations}
For any \textsc{WavePacket} simulation, the structure of the quantum-mechanical Hamiltonian operator must be of the form introduced in Part I 
\begin{equation}
\hat{H}(\hat{R},-i\nabla_R,t)=\hat{T}(\hat{R},-i\nabla_R) + \hat{V}(\hat{R}) - F(t)\cdot\hat{\mu}(\hat{R})
\label{eq:Hamilt}
\end{equation}
where $\hat{T}$ and $\hat{V}$ are kinetic and potential energy operators, respectively. 
They are expressed in terms of position and momentum operators, $\hat{R}$ and $-i\nabla_R$, which can be in one or more dimensions.
The dynamics of the quantum system can be controlled by electrical field components $F_k(t)$ with $k\in\{x,y\}$ allowing to account for different polarization directions. 
Within the semi-classical dipole approximation they interact with the quantum system through its dipole moment components $\mu_k(\hat{R})$.
Note that the induced dipole interaction involving products of the polarizabilities and the field squared has been omitted here because it is not yet implemented in all the codes described below.
The same holds for (optional) negative imaginary potential, which can be used to absorb wavefunctions near the boundaries.
For the description of open quantum systems we employ a simple model of the system--bath coupling (SBC) Hamiltonian. 
Its dependence on the system's degrees of freedom is modeled by functions $\chi(R)$ while the dependence on the bath modes $\hat{r}_\beta$ is assumed to be linear 
\begin{equation}
	\hat{H}_\mathrm{SB}=\sum_\beta \chi_\beta (\hat{R}) \hat{r}_\beta
	\label{eq:sbc}
\end{equation}
where the summation extends over all bath modes $\beta$.
For more details, see Sec.~\ref{sec:qm_abncd} and appendix \ref{sec:appB}.

It is emphasized that \textsc{WavePacket} can also be used to solve coupled Schr\"{o}dinger equations in which case $\hat H$ is an operator matrix and the potential $\hat V$ as well as the dipole moments $\hat{\mu}_k$ become matrix-valued.
This occurs, e.~g., for systems comprising of heavy and light particles, where  $\hat{R}$ and $-i\nabla_R$ refer to the former ones, while the matrices of $\hat V$ and $\hat{\mu}_k$ refer to the quantization of the latter ones.

Within \textsc{WavePacket}, all settings concerning the Hamiltonian (\ref{eq:Hamilt}) have to be specified by the user.
Most conveniently this can be achieved with a user-defined function, which we recommend to name \texttt{qm\_init}.
Typically such a function will also contain further settings, in particular those required for the discrete variable representations (DVRs) an/or the corresponding finite basis representations (FBRs).
Those representations are underlying the functions \texttt{qm\_bound} as well as \texttt{qm\_propa}, both of which are described in great detail in Part I.
However, for completeness, the former one will be shortly reviewed in the following Sec.~\ref{sec:qm_bound}, before switching from DVR / FBR to energy representations in Sec.~\ref{sec:qm_matrix}.

\subsection{Example: Morse oscillator in FFT grid representation}
\label{sec:qm_init:Morse}

We first choose the rather simple example system of a Morse oscillator already introduced in Part I; generalizations to more dimensions and/or more complex scenarios can be found in the demo examples available in the Wiki documentation of \textsc{WavePacket} at \textsc{SourceForge}.
We consider a one-dimensional Morse potential with dissociation energy $D_e$, equilibrium position $R_e$, and range parameter $\alpha>0$, the values of which are chosen to resemble an OH oscillator inside a water molecule in its electronic ground state \cite{BSchmidt:14,Zhu1998,Zhu1998a}.
The Morse system interacts with an external electric field through a dipole moment modeled by a Mecke function \cite{Mecke:50a} with charge and distance parameters $R_0=0.6$ \AA\ and $q_0=7.85$ D/\AA, respectively.  
For the case of open quantum systems, the $R$--dependence of the system-bath coupling, $\chi(R)$, see Sec.~\ref{sec:qm_abncd} and appendix \ref{sec:appB}, is modeled by a linear function with slope one for simplicity.

The necessary settings of the \textsc{MATLAB} function \texttt{qm\_init} could read as follows 
{\small\begin{verbatim}
global atomic hamilt space

hamilt.pot.handle = @pot.morse;
hamilt.pot.d_e = 0.1994; 
hamilt.pot.r_e = 1.821; 
hamilt.pot.alf = 1.189;

hamilt.dip.handle = @dip.mecke;
hamilt.dip.r_0 = 0.6/atomic.l.A;
hamilt.dip.q_0 = 7.85/atomic.d.D*atomic.l.A;

hamilt.sbc.handle = @sbc.taylor;
hamilt.sbc.v{1,1} = 1;

space.dof{1} = grid.fft; 
space.dof{1}.mass = 0.9481/atomic.m.u;
space.dof{1}.n_pts = 256;
space.dof{1}.x_min =  0.7;
space.dof{1}.x_max = 10.0;
\end{verbatim}}

The first line makes the three global variables available inside the function.
Note that it is a general feature of the Matlab version of WavePacket to use few, but highly structured variables to simplify book-keeping and to avoid passing values of arguments between functions as is required in some older versions of Matlab. 

The choices of the Morse potential, the Mecke dipole function, and the linear system bath coupling (Taylor series with first term only) are realized through function handles, i.e., references to functions located within the package folders \texttt{+pot}, \texttt{ +dip}, \texttt{ +sbc}, respectively.
The choice of the FFT--based plane wave FBR is realized by constructing an object pertaining to one of the \textsc{MATLAB} classes in folder \texttt{+grid}. The \texttt{fft} class used here requires the reduced mass, the number of grid points as well as the lower and upper bounds of the grid to be specified.

The use of function handles and grid classes allows easy customization since \textsc{WavePacket} comes with a large number of built-in models, see the Reference Manual on our Wiki pages.
In addition, there is the possibility for the user to supply custom functions and/or classes.
As an alternative, the necessary functions can also be specified in terms of a Taylor series, or they can be given as tabulated values from formatted data files, which are then interpolated.

Throughout the \textsc{WavePacket} software package, atomic units are used, where Planck's constant $\hbar$, the electronic mass $m_e$ and the elementary charge $e$ are scaled to unity. 
However, conversions from and to SI (and other frequently used) units are provided through the fields of global variable \texttt{atomic} as can be seen in some of the sample code lines above.
Note that also the most important isotopic masses of frequently encountered atom types are available there.
In principle, the sample code lines given above are equivalent to those given as an example in Part I. 
However, for a few minor syntax changes, see Appendix \ref{sec:appA}.

\subsection{Workflow}
\label{sec:qm_init:workflow}

Once, the initialization function \texttt{qm\_init} has been set up, a typical workflow for a \textsc{WavePacket} simulation could be as follows
{\small\begin{verbatim}
qm_setup; 
qm_init;
qm_bound; 
qm_cleanup
\end{verbatim}}
where the function qm\_bound for bound state calculations can be exchanged by any of the functions described in the following sections.
For more details of the workflow, as well as the possibilities of more sophisticated \textsc{Matlab} scripts, the reader is referred to  Sec.~2 of Part I.

\section{qm\_bound}
\label{sec:qm_bound}

\subsection{Calculation of bound states}
Once the Hamiltonian, along with the necessary DVR and FBR schemes, is specified in \texttt{qm\_init}, the time-independent Schr\"{o}dinger equation (TISE)
\begin{equation}
\label{eq:tise}
	\hat{H}_0(R,-i\nabla_R)\Psi_j(R)=E_j\Psi_j(R)
\end{equation}
can be solved. Here $E_j$ and $\Psi_j(R)$ are the eigenvalues and eigenfunctions of $\hat{H}_0=\hat{T}+\hat{V}$ which equals the Hamiltonian of Eq.~(\ref{eq:Hamilt}) but without the last (time-dependent) term.
As explained in detail in Part I, the default method to solve the TISE numerically within \textsc{WavePacket} is by direct diagonalization using \texttt{qm\_bound} which is based on DVR and FBR methods. 
The numerical solutions are restricted to the calculation of bound states; calculations of scattering states are planned for future versions of our software package.

In any simulation using \texttt{qm\_bound} (as well as \texttt{qm\_propa} for time-dependent simulations), expectation values of positions, momenta, and energies are routinely monitored.
Moreover, \textsc{WavePacket} offers the possibility to calculate time-independent (or time-dependent) expectation values of additional multiplicative operators (AMOs). 
For example, in simulations of molecular rotation the degree of orientation and/or alignment can be determined as mean values of $\cos\theta$ or $\cos^2\theta$, respectively, where $\theta$ is the angle between the axis of the molecule and the polarization of the external field \cite{BSchmidt:60,BSchmidt:68}.
In simulations of chemical reaction dynamics, projecting on the reactant and/or the product space can be instrumental in monitoring reaction probabilities \cite{Steinfeld:89a}.
So far, these possibilities are restricted to AMO operators which are multiplicative in position representation (DVR).
For future versions, we plan also additional differential operators (ADOs), i.~e., operators which are multiplicative in momentum representation (FBR).

\subsection{Example: Morse oscillator bound states}
\label{sec:qm_bound:Morse}

Here we return to the example of the Morse oscillator and add a few lines to the \textsc{WavePacket} initialization function \texttt{qm\_init} to define an AMO
{\small\begin{verbatim}
space.amo{1}.handle = @amo.gauss;
space.amo{1}.pos_0 = 2.5;
space.amo{1}.width = 1/50;
\end{verbatim}}
Here the function handle in the first code line indicates that we have chosen a Gaussian function which can be used as a target in (optimal) control of molecular bond length \cite{Zhu1998a}, see Sec.~\ref{sec:qm_optimal}.
The corresponding \textsc{Matlab} function \texttt{gauss.m} is located in the \texttt{+amo} package folder in the source code directory, along with a few other model functions.
Again, it is emphasized that such functions can be easily provided by the user, in order to adapt \textsc{WavePacket} to specific simulation requirements.
It is also possible to specify more than one additional multiplicative operator in which case the indices inside the curly braces have to be set appropriately.

Furthermore, it is possible to specify that all 22 bound states of the OH Morse oscillator with $0 \le v \le 21$ are to be visualized and their expectation values to be included in the logfile output which can be achieved by adding the following lines to \texttt{qm\_init}
{\small\begin{verbatim}
global psi
psi.eigen.stop = 0;
psi.eigen.stop = 21;
\end{verbatim}}
Note that a Wigner representation of the highest bound state of our Morse oscillator is shown in Fig.~4 of Part I.

Another recommended option is to store the calculated wave functions which is triggered by two more lines in \texttt{qm\_init}
{\small\begin{verbatim}
psi.save.export = true;
psi.save.file = 'bound'; 
\end{verbatim}}
Here the last line indicates that the wave functions are to be stored in unformatted \textsc{Matlab} data files \texttt{bound.mat, bound\_0.mat, ...}.
These files serve not only as an input to the visualization function \texttt{qm\_movie} described in Sec.~6 of Part I, but they also provide the necessary data for \texttt{qm\_matrix}, see the following section.

\section{qm\_matrix}
\label{sec:qm_matrix}

\subsection{From coordinate to energy representation}
Once the time-independent Schr\"{o}dinger equation (\ref{eq:tise}) has been solved for using DVR / FBR techniques, the function \texttt{qm\_matrix} can be used to set up an energy representation in terms of the obtained eigenenergies, $E_j$, and eigenfunctions, $\Psi_j(R)$, of the field--free Hamiltonian $\hat{H}_0=\hat{T}+\hat{V}$.
The resulting change from PDEs to coupled ODEs makes it easier to set up the equations of motion for closed and open quantum systems on an equal footing, see Sec.~\ref{sec:qm_abncd}.

Within the \textsc{WavePacket} function \texttt{qm\_matrix}, the change from DVR (coordinate) and/or FBR (momentum) to energy (or eigen) representation is achieved by calculating matrix elements of the dipole operator
\begin{equation}
	\mu^{(k)}_{ij} = \langle i | \mu_k | j \rangle = \int {\mathrm d}R \Psi^\ast_i(R) \mu_k(R) \Psi_j(R)
	\label{eq:matrix:dipole}
\end{equation}
where $k\in\{x,y\}$ allows to simulate the interaction with light of different polarization directions.
For open quantum systems, \texttt{qm\_matrix} similarly evaluates matrix elements of the system-bath coupling, $\chi(R)$, see Eq.~(\ref{eq:sbc}) and appendix \ref{sec:appB}.
All integrals are obtained by means of the numerical quadratures underlying the DVRs; for details see Sec.~3.3 of Part I.

Moreover, the function \texttt{qm\_matrix} serves to generate energy representations of observables used as control targets in the (optimal) control functions described in the following sections. 
Currently, there are three  options available: 
\begin{itemize}
	\item Additional multiplicative operators (AMOs) as introduced in Sec.~\ref{sec:qm_bound}. 
	The matrix elements of $\hat{O}_q$ are given by
	\begin{equation}
  O^{(q)}_{ij} = \langle i|\hat{O}_q| j\rangle 
  \label{eq:target:amo}
  \end{equation}
where again the required integrals are calculated by DVR quadratures. 
	\item Populations of (one or more) selected eigenstates. 
	The matrix elements of the corresponding projectors $\hat{P}_q$ are given by
  \begin{equation}
  P^{(q)}_{ij} = \langle i|\hat{P}_q| j\rangle = \langle i|q \rangle \langle q|j \rangle
  \label{eq:target:pop}
  \end{equation}
	in which case the only non-zero matrix elements are ones on the appropriate ($q$-th) position(s) 
	along the main diagonal.
	\item Alternatively, populations can be obtained as squared moduli of overlaps with eigenstates which are simply given as a vector with elements
	\begin{equation}
  \Pi^{(q)}_{i} =  \langle i|q \rangle = \delta_{i,q}
  \label{eq:target:prj}
  \end{equation}
	where the only non-zero elements are ones on the appropriate ($q$-th) position(s).
\end{itemize}
Even though the latter two options are formally equivalent, there are non-trivial differences when using them as targets in numeric optimal control simulations, see Sec.~\ref{sec:qm_optimal:Morse}.

All matrix elements of the dipole operators (\ref{eq:matrix:dipole}), system bath couplings, as well of those of one of the three classes of observables (\ref{eq:target:amo}-\ref{eq:target:prj}) are written to an unformatted \textsc{Matlab} data file named \texttt{tise.mat}. 
Alternatively, these data may also come from other sources outside \textsc{WavePacket}.
For example, in simulations of electronic dynamics of atomic or molecular systems, the necessary matrix elements can be generated by quantum chemical (electronic structure) calculations.

\subsection{Morse oscillator matrix elements}
\label{sec:qm_matrix:Morse}

Here we return to the example of the Morse oscillator already used in the previous sections.
Because its spectrum also contains a continuum of scattering states, the transformation of quantum dynamics  from a DVR/FBR to an energy representation leads to integro-differential equations also accounting for the bound-continuum coupling  \cite{BSchmidt:14}.
However, this coupling can be neglected as long as frequencies and/or intensities of the control fields are not too high, in which case the study of vibrational excitation of a Morse oscillator can still be pursued in an ODE setting.
Otherwise, one would have to resort to numerical techniques for the discretization of a quasi-continuum \cite{Burkey1984} which have been used, e.~g. in modeling dissociation or ionization spectroscopy \cite{Seel1991}.
However, those approaches are currently not yet implemented in \textsc{WavePacket}.

When using the Gaussian-shaped AMO introduced in Sec.~\ref{sec:qm_bound:Morse}  as control target, the following lines have to be added to the \textsc{WavePacket} initialization function \texttt{qm\_init.m}
{\small\begin{verbatim}
global control
control.observe.targets = 'amo';
control.observe.choices = {1};
\end{verbatim}}
If more than one AMO has been specified before, several indices could be given in the cell vector on the r.h.s. of the latter code line.

Alternatively, the choice of bound state populations as control targets is specified as follows
{\small\begin{verbatim}
control.observe.types = 'prj';
control.observe.choices = {[0] [1] [2] [3] [4] [5:10] [11:21]};
\end{verbatim}}
where five single states and two groups of states are given here for illustration. 
The above mentioned possibility of using (squares of) overlaps instead of projectors when using populations as control targets is selected by specifying \texttt{'ovl'} instead of \texttt{'prj'} above.

\section{qm\_abncd}
\label{sec:qm_abncd}

\subsection{From closed to open quantum systems}

The function \texttt{qm\_abncd} is intended to set up simulations of closed and open quantum systems using a common framework in terms of $A,B,N,C,D$ matrices frequently used in the literature on control systems, see Secs.~\ref{sec:qm_abncd:input} and \ref{sec:qm_abncd:output} below.
The evolution of a closed non-relativistic quantum system is described in terms of the time-dependent Schr\"{o}dinger equation (TDSE) 
\begin{equation}
i\frac{d}{dt}|\Psi(t)\rangle = \hat H(\hat{R},-i\nabla_R,t) |\Psi(t)\rangle, \quad 
|\Psi(t=0)\rangle=|\Psi_0\rangle
\label{eq:tdse}
\end{equation}
where the expectation values of the $q$-th observable $\hat{O}_q$ are obtained as mean values
\begin{equation}
  \langle \hat{O}_q\rangle(t) = \langle \Psi(t)| \hat{O}_q | \Psi(t)\rangle 
  \label{eq:mean}
  \end{equation}
While \texttt{qm\_propa} can be used to solve the TDSE (\ref{eq:tdse}) in a PDE setting using DVR/FBR techniques as explained in Part I, the \textsc{WavePacket} function \texttt{qm\_abncd} sets up the ODE formulation in  energy representation, based on the matrix elements obtained from \texttt{qm\_matrix}, see Sec.~\ref{sec:qm_matrix}.

Alternatively, function \texttt{qm\_abncd} can be used to set up simulations of open quantum systems, i.~e., systems thermally coupled to a heat bath \cite{Weiss:99a,May:00a,Breuer:02a}.
Within the Lindblad formalism, the evolution of the reduced density operator $\rho$ is governed by the following quantum master equation (Liouville-von Neumann equation, LvNE)
\begin{equation}
	\frac{d}{dt}\hat{\rho} = -i\left[\hat H,\hat\rho \right]_--\sum_\ell\left\{\hat L_\ell\hat\rho\hat L_\ell^\dagger-\frac{1}{2}\left[\hat L_\ell^\dagger \hat L_\ell,\hat\rho\right]_+\right\}, \quad 
\hat{\rho}(t=0)=\hat{\rho}_0
\label{eq:lvne}
\end{equation}
where $[\cdot,\cdot]_-$ and $[\cdot,\cdot]_+$ stand for commutators and anticommutators, respectively.
The first term of the r.h.s. of (\ref{eq:lvne}) is the Hamiltonian part describing the dynamics of a closed system.
The second term represents the coupling to the environment, i.~e., dissipation and/or dephasing.
For the different choices of Lindblad operators $\hat{L}_\ell$ available inside \textsc{WavePacket}, see appendix \ref{sec:appB}.
Within the LvNE setting, the time dependence of expectation values of observables can be calculated as
\begin{equation}
	\langle\hat{O}_q\rangle(t) = \mathrm{Tr}(\hat{O}_q\hat{\rho}(t)) 
	\label{eq:trace}
\end{equation}
where Tr denotes the trace operation.

\subsection{Input equations}
\label{sec:qm_abncd:input}

In linear time-invariant (LTI) system theory \cite{Zhou1998} the input equation of a control system describes the evolution of its state vector $x(t) \in C^n$
\begin{equation}
\dot{x}(t)=Ax(t)+iBu(t),\,x(0)=x_0
\label{eq:input:linear}
\end{equation}
where the field-free evolution is described by a Hermitian matrix $A \in C^{n\times n}$ with 0 as a  simple eigenvalue and where the interaction with a low-dimensional control, $u(t) \in R^m, m \ll n$, is given by the input matrix $B \in R^{n \times m}$.
However, for quantum systems governed by the Hamiltonian given in (\ref{eq:Hamilt}), driven by external control field(s), $u_k(t) \equiv F_k(t)$, we are dealing with a bi-linear input equation
\begin{equation}
\dot{x}(t)=\left( A+i\sum_{k=1}^m u_k(t)N_k\right)x(t),\quad x(0)=x_0
\label{eq:input:bilinear}
\end{equation}
where now the control term depends on both the field components, $u_k(t)$, and the state vector, $x(t)$. 
While identification with the TDSE (\ref{eq:tdse}) is straightforward, re-writing the LvNE (\ref{eq:lvne}) into this form is based on suitable vectorization of the density matrix $\rho(t)$.
For the corresponding matrix representations of the commutators and anticommutators, see appendix A of Ref.~\cite{BSchmidt:61}.

Next, we introduce an equilibrium state $x_e$ defined as $A x_e = 0$.
In the case of the TDSE~(\ref{eq:tdse}), this is typically the ground state (after shifting its energy to zero), whereas in case of the LvNE~(\ref{eq:lvne}) this is the thermal equilibrium defined by temperature $\Theta$ in the construction of the Lindblad operators obeying microscopic reversibility (\ref{eq:balance}), see App.~\ref{sec:appB}.
Upon shifting the state vectors $x(t) \rightarrow x(t) - x_e$ and setting $b_k \equiv N_k x_e$, the following equation of motion is retrieved
\begin{equation}
\dot{x}(t)=\left( A+i\sum_{k=1}^m u_k(t)N_k\right)x(t)+i\sum_{k=1}^m u_k(t)b_k,\quad x(0)=x_0-x_e
\label{eq:input:shifted}
\end{equation}
which is implemented in the \textsc{WavePacket} functions \texttt{qm\_control} and \texttt{qm\_optimal} described in Secs.~\ref{sec:qm_control} and \ref{sec:qm_optimal}. 
The shifted equation is now inhomogeneous and therefore more complicated. 
However, for an equilibrium initial condition, we have $x(0) = 0$ which is required for dimension reduction, see Secs.~\ref{sec:qm_balance}-\ref{sec:qm_H2model}.

\subsection{Output equations}
\label{sec:qm_abncd:output}

In LTI system theory \cite{Zhou1998}, the output equation defines observables $y(t) \in R^p, p\ll n$, in terms of an output matrix $C \in R^{p\times n}$
\begin{equation}
y(t)=Cx(t)
\end{equation}
For open quantum systems described by the LvNE (\ref{eq:lvne}), we rewrite this in terms of components 
\begin{equation}
y_q(t)=c_q^\dagger x(t)+c_q^\dagger x_e,\quad q=1,\ldots,p
\label{eq:output:linear}
\end{equation}
where we have again shifted the state vectors $x(t) \rightarrow x(t) - x_e$ and where the observables are represented by vectors $c_q \in R^n$.
They are obtained by suitable mapping of the trace formula (\ref{eq:trace}), see again appendix A of Ref.~\cite{BSchmidt:61}.

For closed quantum systems described by the TDSE (\ref{eq:tdse}), one has to consider quadratic output equations 
\begin{equation}
y_q(t)=x^\dagger(t)D_qx(t)+2\Re(x_e^\dagger D_qx(t))+x_e^\dagger D_qx_e,\quad q=1,\ldots,p
\label{eq:output:quadratic}
\end{equation}
where every observable is represented by a Hermitian matrix $D_q \in R^{n\times n}$ obtained as a matrix representation of Eq. (\ref{eq:mean}).
Because the \textsc{WavePacket} functions \texttt{qm\_control} and \texttt{qm\_optimal} offer the choice of linear (\ref{eq:output:linear}) or quadratic (\ref{eq:output:quadratic}) output, those functions can be used both for LvNE and TDSE control problems.

\subsection{Usage notes}
\label{sec:qm_abncd:usage}

By default, the \textsc{WavePacket} function \texttt{qm\_abncd} reads data from unformatted \textsc{Matlab} data file \texttt{tise.mat} provided by function \texttt{qm\_matrix}, see Sec.~\ref{sec:qm_matrix} above. 
After having set up the $A,B,N,C$, or $D$ matrices, the \textsc{WavePacket} function \texttt{qm\_abncd} writes them to unformatted data files named \texttt{tdse.mat} or \texttt{lvne.mat} for simulations of closed or open quantum systems, respectively.
Alternatively, data files containing $A,B,N,C$, or $D$ matrices can also come from other sources.
As an example we mention here semi-discretized Fokker-Planck equations, which can also be written in the form of Eqs.~(\ref{eq:input:bilinear}) or (\ref{eq:input:shifted}), see our work in Refs.~\cite{hartmann2010e,BSchmidt:77}.

\subsection{Example: Morse oscillator with dissipation}
\label{sec:qm_abncd:Morse}

As an example, let us consider the Morse oscillator from the previous sections, now interacting with a thermal bath through the linear SBC model of Eq.~(\ref{eq:sbc}).
This can be realized by adding the following lines to the \textsc{WavePacket} initialization function \texttt{qm\_init} before running \texttt{qm\_abncd('lvne')}
\begin{verbatim}
global control
control.lvne.temperature = 0;
control.lvne.order = 'df';
control.relax.rate = 2*atomic.t.ps;
control.relax.lower = 0;
control.relax.upper = 1;
control.relax.model = 'fermi';
\end{verbatim}
The second line is used to set the temperature, here $\Theta=0$, while the third line specifies the ordering of the density matrix elements. Here '\texttt{df}' stands for "diagonals first" which is the recommended option, see again App.~A of Ref.~\cite{BSchmidt:61}.
The next three lines serve to set the relaxation rate, here $\Gamma_{0\leftarrow 1}=2$ ps$^{-1}$, and the last line is intended to select the relaxation model (based on Fermi's golden rule) to calculate all other rates, as described in App.~\ref{sec:appB}.

\begin{figure}
	\caption{Spectrum of the $A$ matrix for the Morse oscillator of Ref.~\cite{BSchmidt:14} (bound states only). The coupling to a thermal bath is described by a Lindblad model, using Fermi's golden rule (\ref{eq:fermi}) for the relaxation rates. The matrix is generated using the \textsc{WavePacket} function \texttt{qm\_abncd}, assuming a relaxation rate $\Gamma_{0\leftarrow 1}=2$ ps$^{-1}$ and temperature $\Theta=0$.}
	\includegraphics[width=10cm]{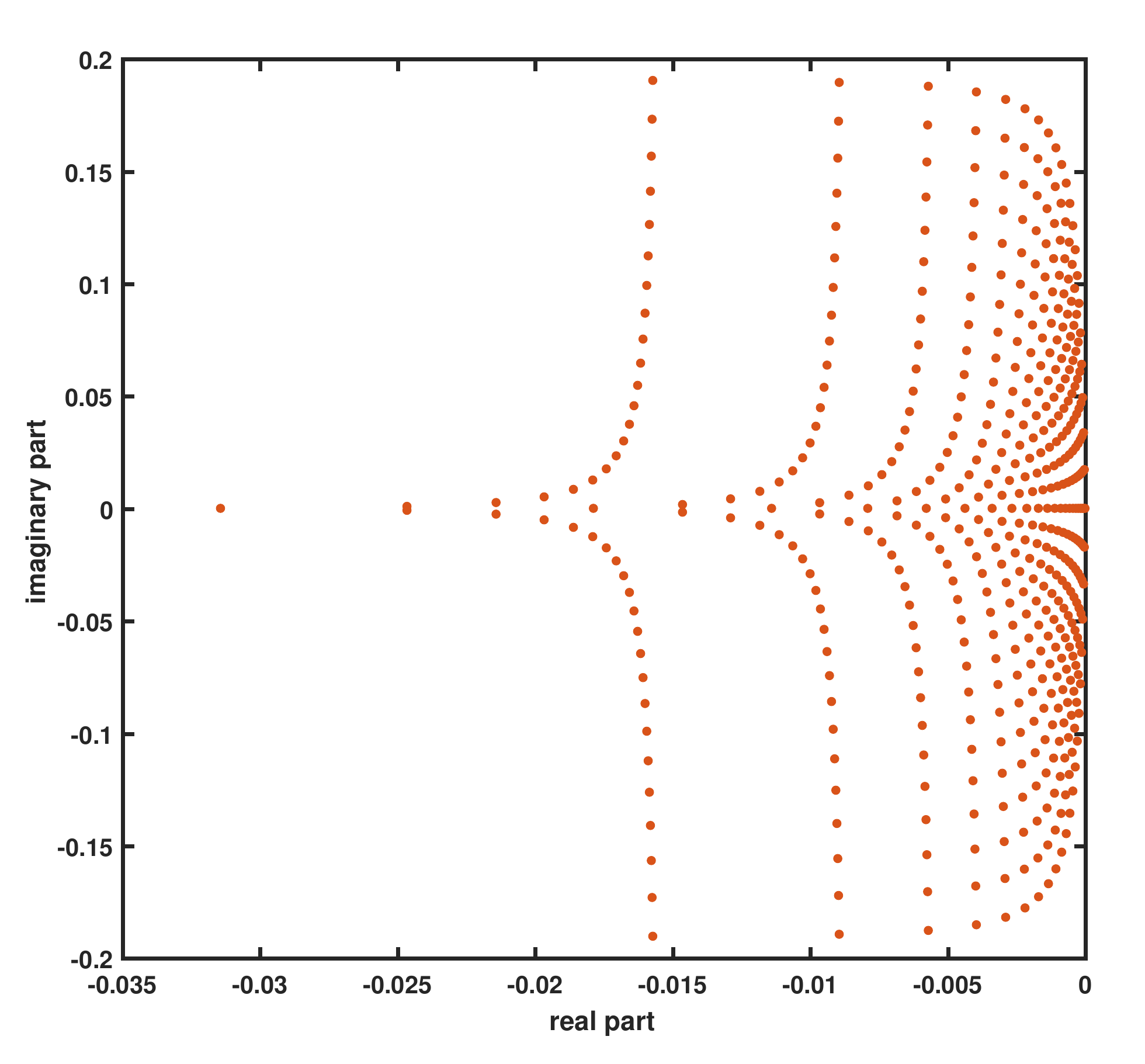}
	\label{fig:qm_abncd}
\end{figure}

Here, the resulting matrices $A$ and $N$ are of dimension $484 \times 484$ with a density of 0.3 \% and 8.8 \%, respectively.
Hence, our codes are exploiting the \textsc{Matlab} support for sparse matrices.
As an example we show the spectrum of matrix $A$ in Fig.~\ref{fig:qm_abncd}. 
The imaginary parts of the eigenvalues give the Bohr frequencies for transitions between bound states of the Morse oscillator under investigation whereas the real parts are essentially determined by the total dephasing rates \cite{BSchmidt:61}.
Note that more negative values of the real parts indicate faster decay. 
We observe that the dephasing is fastest for transitions between states which are energetically near-by, see Eq.~(\ref{eq:fermi}) and Ref.~\cite{Andrianov2006}.

Furthermore, the function \texttt{qm\_abncd} also serves to specify the initial quantum state $|\Psi_0\rangle$ or the initial density matrix $\hat{\rho}_0$ for solving the TDSE (\ref{eq:tdse}) or the LvNE (\ref{eq:lvne}), respectively.
Here we select a pure $v=5$ state which can be prepared, e.~g., employing an intense 1 ps infrared laser pulse, see Ref.~\cite{BSchmidt:14} as well as Sec.~4.5 of Part~I. 
\begin{verbatim}
control.initial.choice = 'pure'; 
control.initial.pure = 5;
\end{verbatim}
Other possible keywords for initializing an LvNE simulation of open quantum systems are \texttt{'cat'} or \texttt{'mixed'} for a coherent ("Schr\"{o}dinger cat") or incoherent superposition of two states, respectively, or \texttt{'thermal'} for a Boltzmann distribution.

Finally, in function \texttt{qm\_abncd} there is also a choice of which of the observables defined in \texttt{qm\_matrix} should be used as control targets.
For example, setting the following
\begin{verbatim}
control.observe.targets = 1:7;
\end{verbatim}
serves to specify that all 7 observables (of type \texttt{'prj'}) defined near the end of Sec.~\ref{sec:qm_matrix:Morse} are to be used for the output equations.

\section{qm\_control}
\label{sec:qm_control}

\subsection{Solving the bi-linear control system}
After the TDSE~(\ref{eq:tdse}) or the LvNE~(\ref{eq:lvne}) describing the dynamics of a closed or open quantum system, respectively, have been cast into the system matrices $A, B, N, C$ or $D$ by virtue of the \textsc{WavePacket} function \texttt{qm\_abncd}, the function \texttt{qm\_control} can be used to solve the corresponding bi-linear control problem.
It consists of the input equation (\ref{eq:input:shifted}) and output equations (\ref{eq:output:linear}) or (\ref{eq:output:quadratic}), see  Secs.~\ref{sec:qm_abncd:input} and \ref{sec:qm_abncd:output}.
Normally, function \texttt{qm\_control} solves the input equation by means of one of \textsc{Matlab}'s built-in ODE solvers.
By default, it uses \texttt{ode45}, a versatile medium order method for systems of non-stiff ODEs; other choices are also possible.
In doing so, each of the main time steps specified by the user, see Sec.~4.4 of Part~I, is adaptively divided into sub-steps using a Dormand-Prince scheme, with the number of sub-steps depending on the relative tolerance which can be specified by the user.
In addition to writing all relevant output to data files, function \texttt{qm\_control} also generates graphics, in particular showing the control field $u(t)$, the state vector $x(t)$, and the observable output $y(t)$.

\subsection{Example: Vibrational control and relaxation in a Morse oscillator}
\label{sec:qm_control:Morse}

\begin{figure}
	\caption{Field-free population dynamics for the Morse oscillator of Ref.~\cite{BSchmidt:14} during the first picosecond (ps). The coupling to a thermal bath is described by a Lindblad model, using Fermi's golden rule (\ref{eq:fermi}) with relaxation rate $\Gamma_{0\leftarrow 1}=2$ ps$^{-1}$ and temperature $\Theta=0$. The evolution is simulated using the \textsc{WavePacket} function \texttt{qm\_control}, assuming the system initially to be in the $v=5$ state. Note that 1 ps corresponds to 41,341 atomic units of time.}
	\includegraphics[width=10cm]{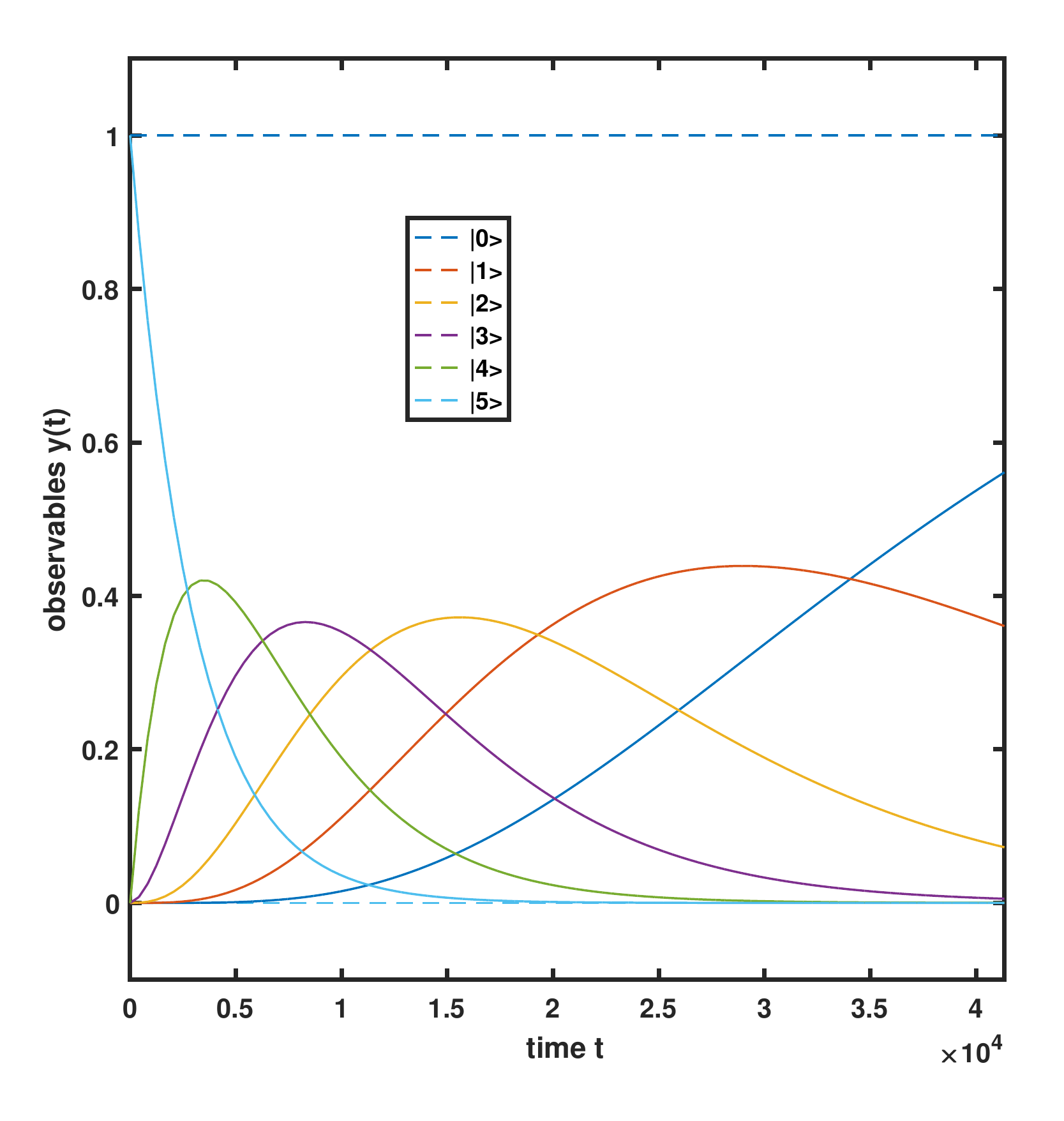}
	\label{fig:qm_control_1}
\end{figure}

Here we continue with our discussion of the Morse oscillator of Ref.~\cite{BSchmidt:14} with dissipation as detailed in Sec.~\ref{sec:qm_abncd:Morse}, again using the Lindblad model of Eq.~(\ref{eq:fermi}) with relaxation rate $\Gamma_{0\leftarrow 1}=2$ ps$^{-1}$.
In order to simulate field-free relaxation dynamics occurring during 1 ps, divided into 100 main time steps, the following lines are added to \texttt{qm\_init}

\begin{verbatim}
global time
time.main.delta = 10/atomic.t.fs;
time.main.start = 0;
time.main.stop = 100;

control.solvers.handle2 = @ode45;
control.solvers.reltol = 1e-6; 
\end{verbatim}
where the last two lines serve to select the function handle for \textsc{Matlab}'s Dormand-Prince \texttt{ode45} integrator and to specify the relative tolerance of the numerical integration.
Fig.~\ref{fig:qm_control_1} shows the vibrational relaxation dynamics for the chosen Morse oscillator. 
Assuming only the $v=5$ state to be initially prepared, the system relaxes in a ladder-wise fashion until the ground state population takes over around $t=824$ fs. 

\begin{figure}
	\caption{Field-induced population dynamics for the Morse oscillator of Ref.~\cite{BSchmidt:14}. The vibrational excitation with a strong infrared laser pulse during the first picosecond (ps), same as in Fig.~2 of part~I, competes with the relaxation, same as in Fig.~\ref{fig:qm_control_1} here. The evolution is simulated using the \textsc{WavePacket} function \texttt{qm\_control}, assuming the system initially to be in the $v=0$ state. Note that 1 ps corresponds to 41,341 atomic units of time.}
	\includegraphics[width=10cm]{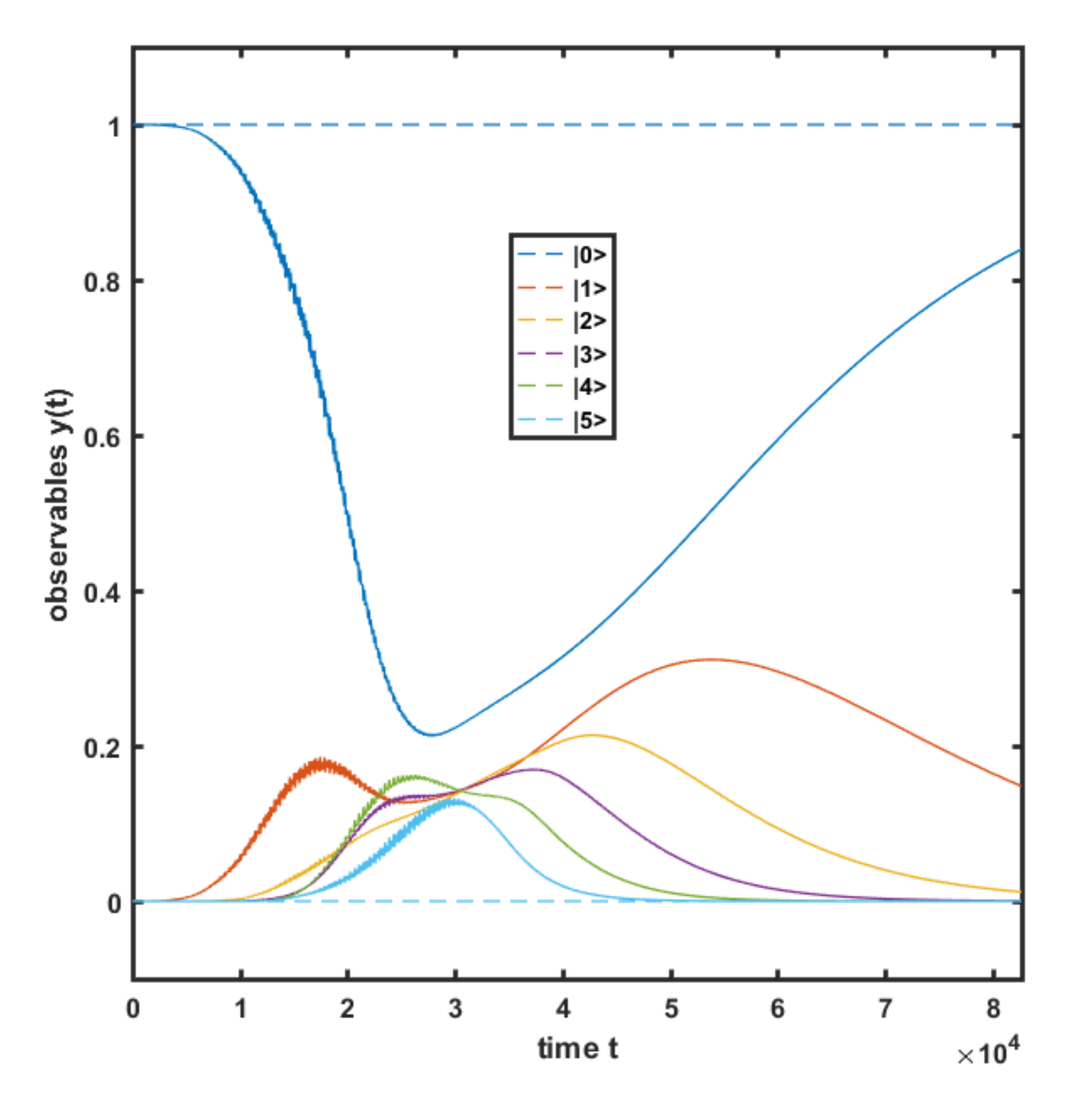}
	\label{fig:qm_control_2}
\end{figure}

Also the competition of vibrational excitation and relaxation can be studied using the function \texttt{qm\_control}.
Fig.~\ref{fig:qm_control_2} shows the results of a 2 ps simulation using the same relaxation rates as before but employing a strong, intense, infrared laser pulse during the first picosecond, see Sec.~4.5 of Part~I. 
As explained there, the pulse is designed to transfer nearly 100\% of the population from the $v=0$ to the $v=5$ state for the case of a closed quantum system ($\Gamma\rightarrow 0$).
Here, however, the coupling to the environment reduces this population transfer down to 13\%. 
At the same time, the vibrational state selectivity is completely lost as can be seen from the approximately equal populations of $1\le v \le 5$ states around $t=726$ fs.

\section{qm\_optimal}
\label{sec:qm_optimal}

\subsection{Quantum optimal control theory}
\label{sec:qm_optimal:theory} 

This section deals with the application of optimal control theory (OCT) to a bi-linear control system consisting of input equation (\ref{eq:input:shifted}) and output equations (\ref{eq:output:linear}) or (\ref{eq:output:quadratic}), as described in Secs.~\ref{sec:qm_abncd} and \ref{sec:qm_control}. 
In the most frequently used version of OCT in quantum dynamics, the final time $T$ is fixed and the task is to find field(s) $u_k(t)$ that drive the system from its initial state $x(t=0)=x_0$ to a final state $x(t=T)$ such that a specified observable $\kappa$ is maximized at final time. 
This is equivalent to maximizing one of the three forms of linear or quadratic target functionals implemented in \textsc{WavePacket}
\begin{eqnarray}
J_{1a}[u,x]&=&x^\dagger (T)D_\kappa x(T)+2\Re(x_e^\dagger D_\kappa x(T)) \nonumber \\
J_{1b}[u,x]&=&\Re\left(c_\kappa^\dagger x(T)\right) \nonumber \\
J_{1c}[u,x]&=&\left|c_\kappa^\dagger x(T)\right|^2 \label{eq:target} 
\end{eqnarray}
where constant terms $c_\kappa^\dagger x_e$ or $x_e^\dagger D_\kappa x_e$ resulting from the equilibrium shift of the input equation (\ref{eq:input:shifted}) have been omitted. 
Functionals $J_{1a}$ and $J_{1b}$ are for optimization of the expectation value of a (positive definite) operator in simulations of a closed (TDSE, \cite{Zhu1998a}) or open (LvNE, \cite{Ohtsuki1999}) quantum system, respectively, see also Eqs.~(\ref{eq:output:quadratic}) and (\ref{eq:output:linear}).
Note again that in the latter case the vector $c_\kappa$ is obtained by suitable vectorization of matrix $D_\kappa$ representing the target observable, see Sec.~\ref{sec:qm_abncd:output}.
The third functional $J_{1c}$ is for the special case of obtaining populations from overlaps in TDSE simulations, see our remarks in Sec.~\ref{sec:qm_matrix} and Ref.~\cite{Zhu1998}.
Even though we assume here the optimization of a single target observable only, generalization to multi-target OCT is straight-forward \cite{Vivie:07a,Zhu2013}.

In addition to maximizing the target functional, it is often of importance to keep the energy associated with the control field (e.~g., the laser fluence) bounded.
Formally, this requirement can be expressed in terms of a cost functional
\begin{equation}
J_2[u]=\sum_k\alpha_k\int_0^Tdt\,u_k^2 (t)/s_k(t)
\label{eq:cost}
\end{equation}
which is to be minimized.
The penalty factors $\alpha_k>0$ balance the importance of the cost functional against that of the target functional and/or balance the importance among the various field components $k$. 
The shape functions $s_k(t)$ can be used to enforce certain shape(s) of the field envelope, e.~g., to model the typical switch on/off behavior of pulsed control fields \cite{Sundermann:99a}. 

Finally, the requirement of physically correct evolution of the system can be written as another functional to be minimized
\begin{equation}
J_3[u,x,z]=2\Re\int_0^Tdt\,z^\dagger(t)(\partial_t-\hat{L}(t))(x(t)+x_e)
\label{eq:constraint}
\end{equation}
where a Lagrange multiplier $z(t)$ has been introduced here to enforce that the state vector $x(t)$ satisfies its evolution equation $\partial_t x(t) = \hat{L}(t) (x(t)+x_e)$ and where the operator $\hat{L}$ stands for the right-hand side of Eq.~(\ref{eq:input:shifted}). 
Since we require the evolution to be fulfilled by the Hermitian conjugate of the evolution as well, we consider here the real part of the functional. 
We note that the inclusion of further constraints is also possible, see for example Refs.~ \cite{Ohtsuki2004,Werschnik2007} which is, however, not yet implemented in \textsc{WavePacket}.

The necessary conditions for the combined functional $J\equiv J_1-J_2-J_3$ to become extremal can be seen directly from Pontryagin's principle. 
For a detailed derivation by means of standard variational calculus we recommend the tutorial by Werschnik and Gross \cite{Werschnik2007}.
The three conditions are as follows:
\begin{enumerate}
\item
The state vector $x(t)$ is propagated forward via the input equation (\ref{eq:input:shifted})
\begin{equation}
\dot{x}(t)=\hat{L} x(t)=\left(A+i\sum_{k=1}^mu_k(t)N_k\right)\left(x(t)+x_e\right)
\label{eq:forward}
\end{equation}
starting from the initial condition $x(t=0) = x_0$.
\item
The Lagrange multiplier $z(t)$ is propagated backward via the adjoint equation
\begin{equation}
\dot{z}(t)=-\hat{L}^\dagger z(t)=\left(-A^\dagger+i\sum_{k=1}^mu_k(t)N^\dagger_k\right)z(t)
\label{eq:backward}
\end{equation}
starting from the final condition $z(t=T)=D_\kappa x(T) + D_\kappa x_e$ when optimizing $J_{1a}$ or from $p(t=T)=c_\kappa$ when optimizing $J_{1b}$ or $J_{1c}$.
For the special case of anti-Hermitian evolution $\hat{L}$ (e.~g. anti-Hermitian $A$ and real symmetric $N$ for closed quantum systems, TDSE) we have the same propagators for state vector $x(t)$ and the Lagrange multiplier $z(t)$. 
\item
The optimized control field for control targets of the form $J_{1a}$ or $J_{1b}$, see Eq.~(\ref{eq:target}), is given by
\begin{equation}
u_k(t)=-\frac{s_k(t)}{\alpha_k}\Im\left(z^\dagger(t) N_k\left(x(t)+x_e\right)\right)
\label{eq:optimal}
\end{equation}
Optimizing targets of type $J_{1c}$ in Eq.~(\ref{eq:target}), as proposed in Ref. \cite{Zhu1998} for the TDSE case and in Ref. \cite{Ohtsuki1999} for the LvNE case, offers the advantage that the initial condition of the backward propagations becomes independent of previous (forward) propagations. 
However, in such cases there has to be an additional factor $\left(x(t)+x_e\right)^\dagger z(t)$ inside the imaginary part of Eq.~(\ref{eq:optimal}) for the optimal control field.
While in theory this factor is independent of time $t$, in the practice of numerical optimization this is often not the case. 
Some authors, cf., Ref. \cite{Werschnik2007} evaluate this at every time $t$, others recommend choosing $t=0$ for forward and $t=T$ for backward propagations, see e.~g. the Appendix of Ref.~\cite{Zhu1998}.
In \textsc{WavePacket} there is a choice between all three options.
\end{enumerate}

Within the \textsc{WavePacket} function \texttt{qm\_optimal} there is a choice of different Runge--Kutta and related integrators to solve the first order ODEs giving the evolution (\ref{eq:forward}) of state vector $x(t)$ and the evolution (\ref{eq:backward}) of Lagrange multiplier $z(t)$ which are coupled through the optimal control field(s) $u_k(t)$.
Typically, these integrators also require knowledge about the value of the field at different times within each of the discretization interval, e.~g. $u(t+\Delta t/2)$ for evaluating the midpoint rule.
As suggested in Refs.~\cite{Zhu1998,Zhu1998a}, the time dependence of $u_k(t)$ can be approximated by a linearization.
The necessary derivative of the field are readily obtained by inserting (\ref{eq:forward}) and (\ref{eq:backward}) into the derivative of (\ref{eq:optimal})
\begin{equation}
\dot{u}_k(t)=-\frac{s_k(t)}{\alpha_k}\Im\left(z^\dagger(t)(N_kA-AN_k)\left(x(t)+x_e\right)\right)
\label{eq:optimal:d}
\end{equation}
Note that within the \textsc{WavePacket} function \texttt{qm\_optimal} only integrators with fixed substep size can be used; for example several Runge-Kutta type methods have been implemented.

\subsection{Iterative schemes}
\label{sec:qm_optimal:iterative} 

The system of the three coupled control equations (\ref{eq:forward})--(\ref{eq:optimal}) is routinely solved by the rapid monotonically convergent iterative algorithms of Refs.~\cite{Zhu1998,Zhu1998a,Ohtsuki1999,maday2003,Ohtsuki2004,Werschnik2007}.
These schemes are initialized by propagating the state vector $x(t)$ forward in time using Eq.~(\ref{eq:forward}).
In doing so, the initially ("guess") field is typically assumed to be constant in time, and its amplitude has to be chosen strong enough to induce some notable dynamics.
Then each step (for $n \ge 1$) of the iteration consists of the following:
\begin{itemize}
\item Propagate the Lagrange multiplier $z(t)$ backward in time using Eq.~(\ref{eq:backward}) with the field
\begin{equation}	
\bar{u}_k^{(n)}(t)=(1-\eta)u_k^{(n-1)}(t)-\eta\frac{s_k(t)}{\alpha_k}\Im\left(\left(z^{(n)}(t)\right)^\dagger N_k\left(x^{(n-1)}(t)+x_e\right)\right)
\label{eq:optimal:b}
\end{equation}
\item Propagate the state vector $x(t)$ forward in time using Eq.~(\ref{eq:forward}) with the field 
\begin{equation}	
    u_k^{(n)}(t)=(1-\zeta)\bar{u}_k^{(n)}(t)-\zeta\frac{s_k(t)}{\alpha_k}\Im\left(\left(z^{(n)}(t)\right)^\dagger N_k\left(x^{(n)}(t)+x_e\right)\right)
		\label{eq:optimal:f}
\end{equation}
\end{itemize}
This is repeated until the change in the total functional $J^{(n)}-J^{(n-1)}$ falls below a user-specified threshold.
The two coefficients $\eta$ and $\zeta$ describe the mixing of fields obtained for the recent and the previous iteration steps. 
In Refs.~\cite{maday2003,LeBris2003} is is shown that monotonic convergence is found if $\eta$ and $\zeta$ are between 0 and 2. 
Note that for the special case of $\eta=\zeta=1$, we retrieve the scheme introduced by Zhu, Botina, and Rabitz \cite{Zhu1998,Zhu1998a}, while for $\eta=0$ and $\zeta=1$ the Krotov method is retrieved \cite{somloi1993}. 
For a specific LvNE example system, the convergence behavior for different values of $\eta$ and $\zeta$ is investigated numerically in Ref.~\cite{Ohtsuki2004}, showing large variations in the number of iteration steps required to achieve a specified tolerance. 

The above algorithms are frequently used in the molecular physics/chemistry community where molecular states are typically manipulated by pulsed lasers.
However, in the physics community, often dealing with the manipulation of spin systems by NMR, there appears to be a preference for gradient ascent pulse engineering (GRAPE) algorithms \cite{Khaneja2005,Machnes2011}.
Those will be included in future versions of \textsc{WavePacket}, with the aim of allowing for direct numeric comparisons for our default demonstration examples.

Finally, it is noted that the function \texttt{qm\_optimal} also generates  graphics 'on the fly', i.~e., the control field $u(t)$, the state vector $x(t)$, and the observable output $y(t)$ can be viewed during the repeated forward and backward propagation.
The animated graphics is also stored as an MPG file for later use in presentations etc.

\subsection{Example: Optimized population transfer in a Morse oscillator}
\label{sec:qm_optimal:Morse}
We return to the example of the Morse oscillator used throughout the previous sections.
Here we consider the fundamental excitation from the ground state $|0\rangle$ to the first excited state  $|1\rangle$.
Even though this is a rather simple task, it shall serve here to illustrate the use of \textsc{WavePacket} for such a quantum optimal control simulation.
The following keywords have to be set within \texttt{qm\_init.m} before running \texttt{qm\_optimal}
\begin{verbatim}
control.optimal.terminal = 2;
control.optimal.max_iter = 50;
control.optimal.tolerance = 1e-10;
control.optimal.alpha = 1.0;
control.optimal.eta  = 1.0;
control.optimal.zeta = 1.0;
\end{verbatim}
The first line specifies the control target (to be evaluated at terminal time), here the second of the seven different populations specified at the end of Sec.~\ref{sec:qm_matrix:Morse}, i.~e. the population of the first vibrationally excited state of the Morse oscillator.
The following two lines serve to set the termination criteria for the iterative procedure, either after 50 iterations or after the change in the overall control functional falls below the specified tolerance. 
The remaining lines specify the "penalty" factor $\alpha$ as well as the prefactors $\eta$ and $\zeta$, see Eqs.~(\ref{eq:optimal:b},\ref{eq:optimal:f}). 

The initial guess for the control field is set in the same way as for the \textsc{WavePacket} functions q\texttt{m\_propa} and \texttt{qm\_control}; for an example see Sec.~4.5 in Part~I.
Note that the shape of the envelope, here a $\sin^2$-shaped half wave of 1 ps duration, serves also as a shape function $s(t)$ during the simulations, see Eq.~(\ref{eq:optimal}).

\begin{figure}
	\caption{Optimal control of population for the Morse oscillator of Ref.~\cite{BSchmidt:14}, using the \textsc{WavePacket} function \texttt{qm\_optimal}. From left to right: TDSE simulation for functional $J_{1a}$, LvNE for $J_{1b}$, TDSE for $J_{1c}$. Top to bottom: Control field $u(t)$, population dynamics $y(t)$, target functional $J_1(t)$, cost functional $J_2(t)$.}
		\includegraphics[width=5.4cm]{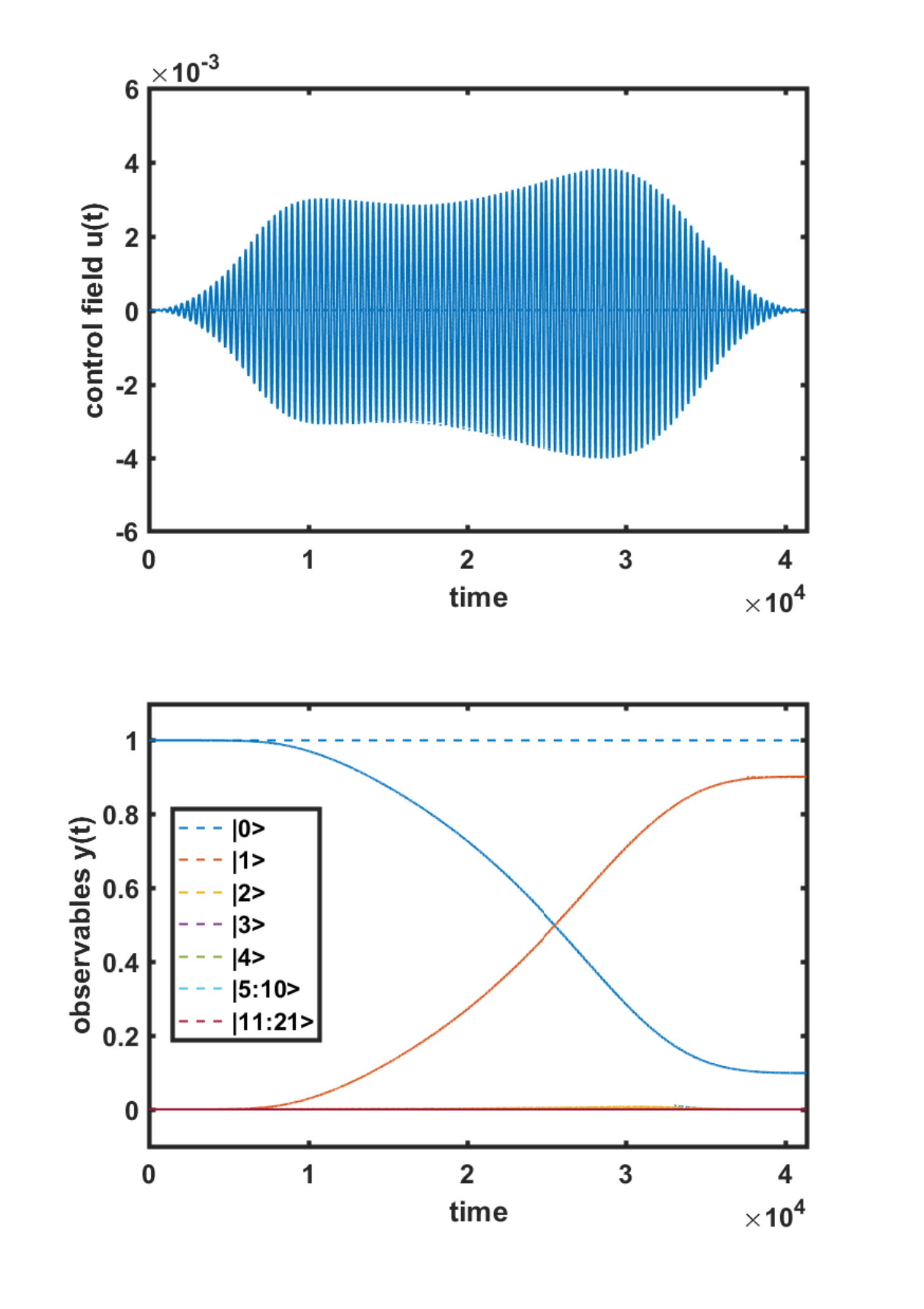}
		\includegraphics[width=5.4cm]{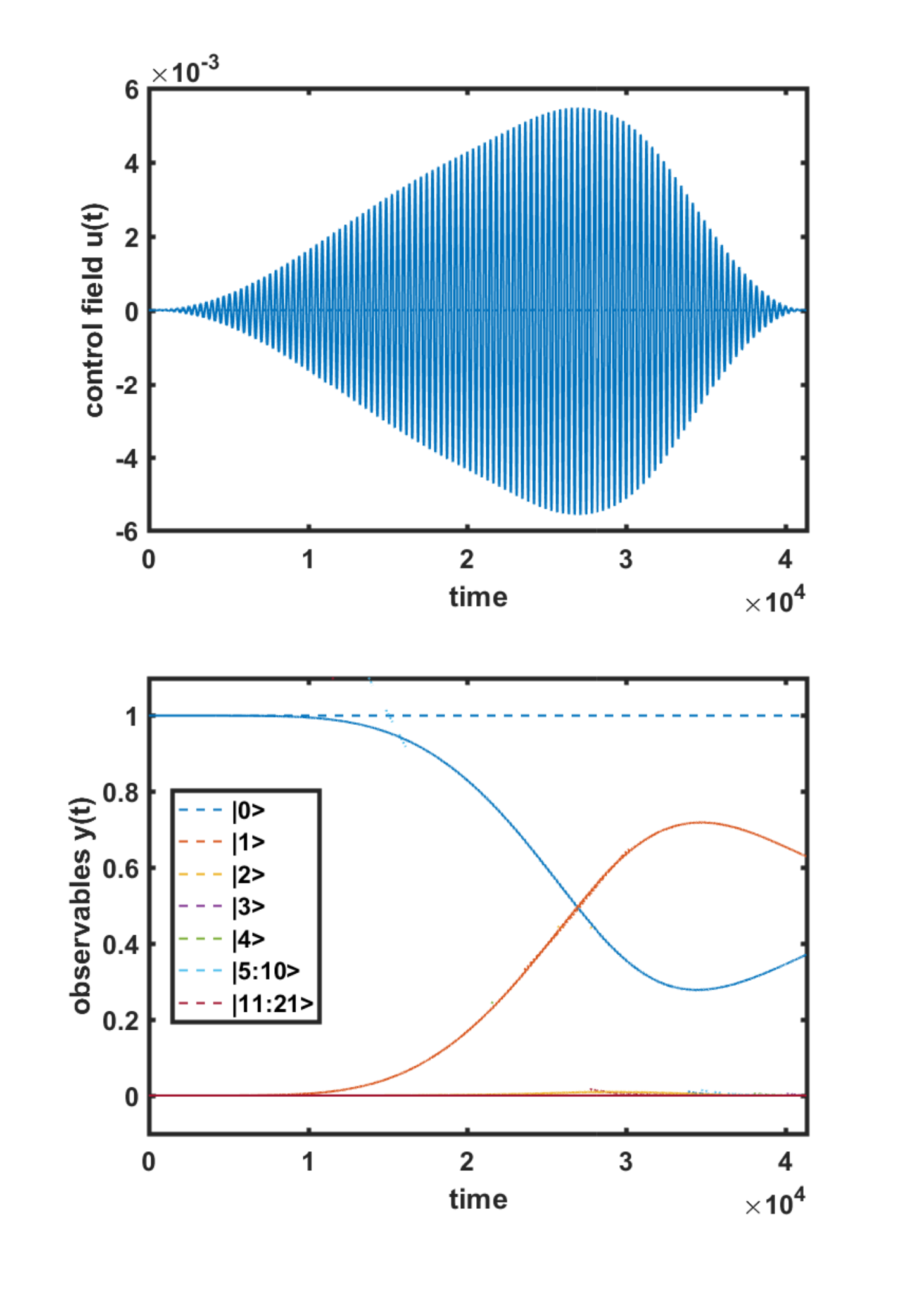}
		\includegraphics[width=5.4cm]{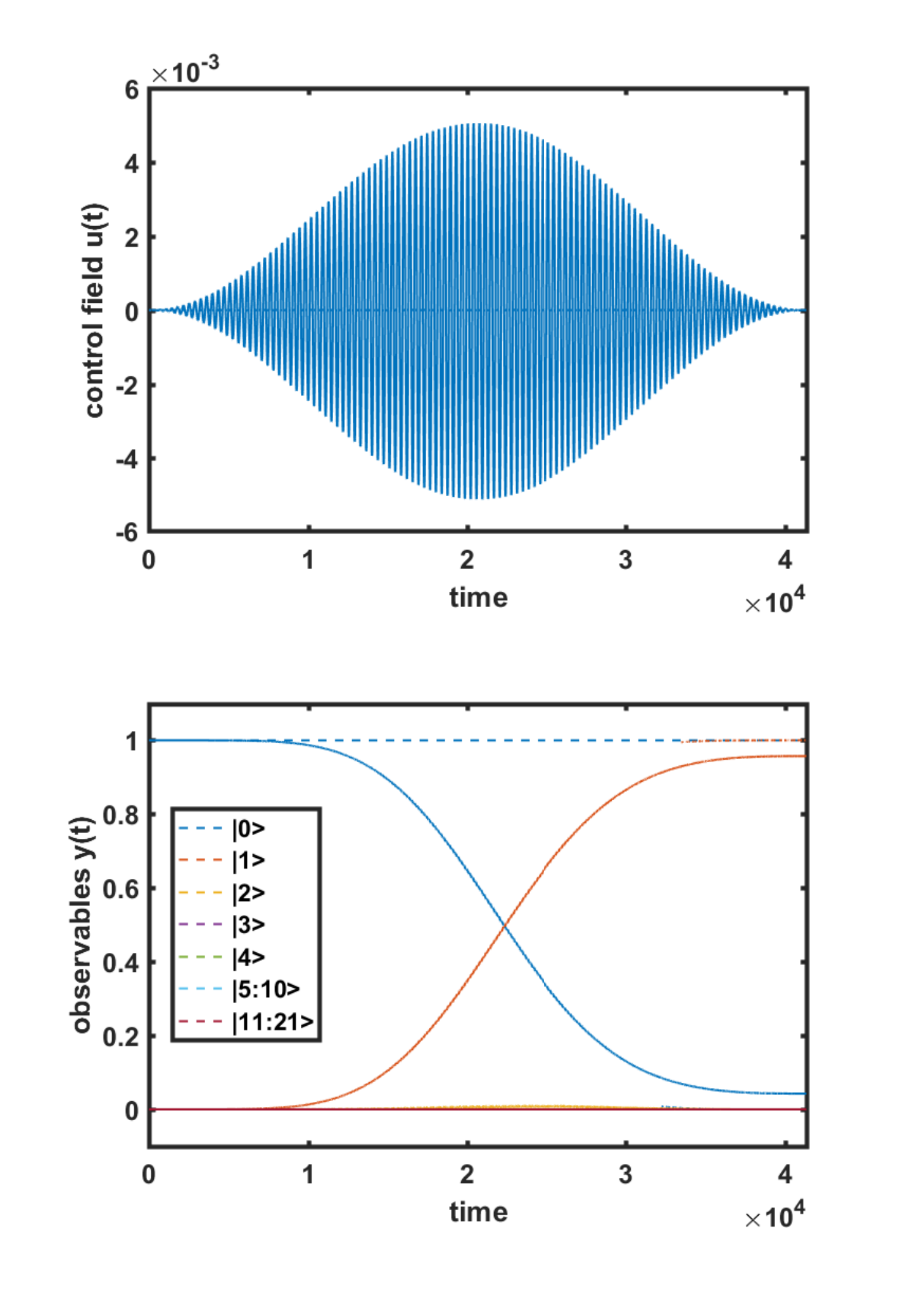}
		\includegraphics[width=5.4cm]{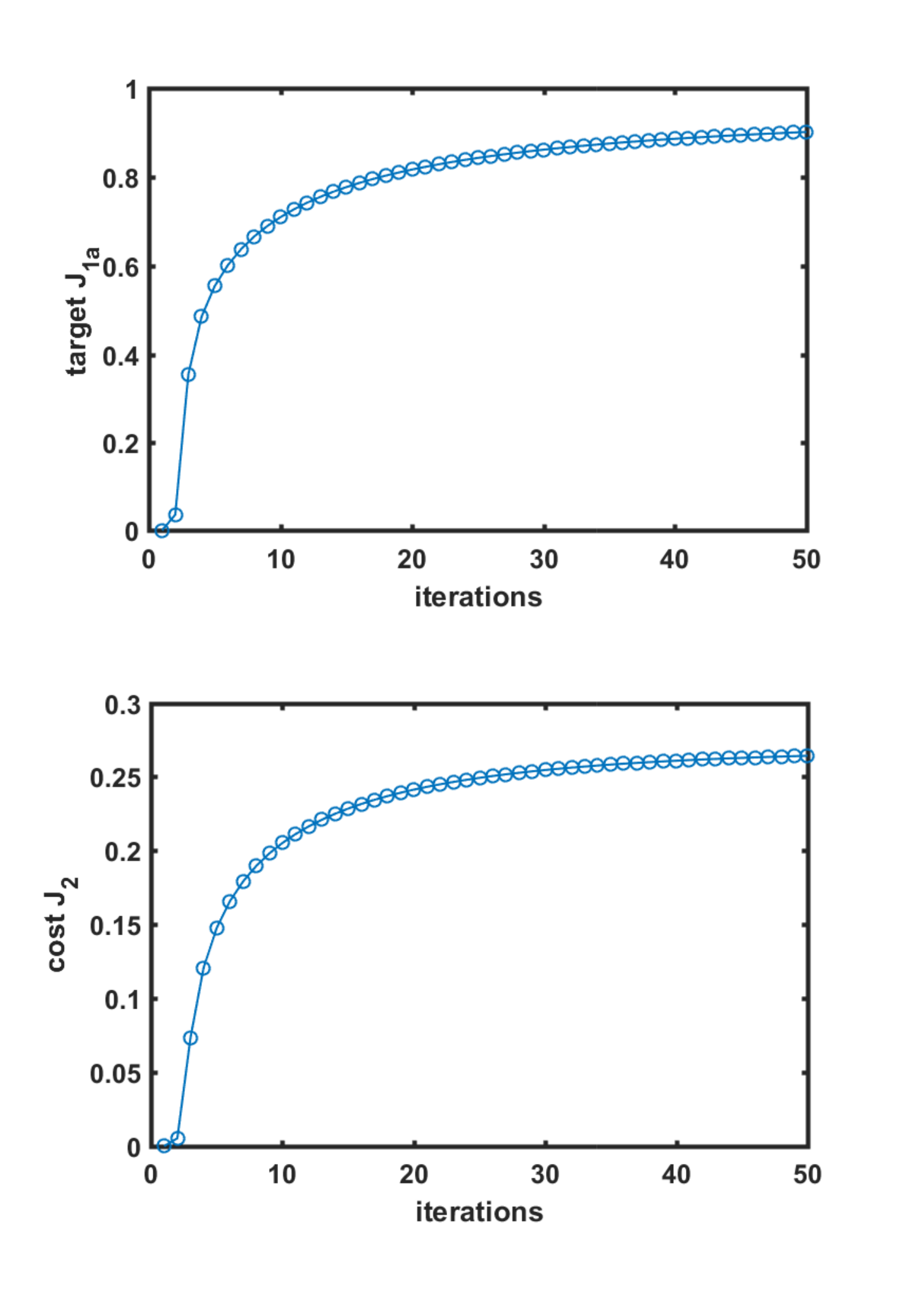}
		\includegraphics[width=5.4cm]{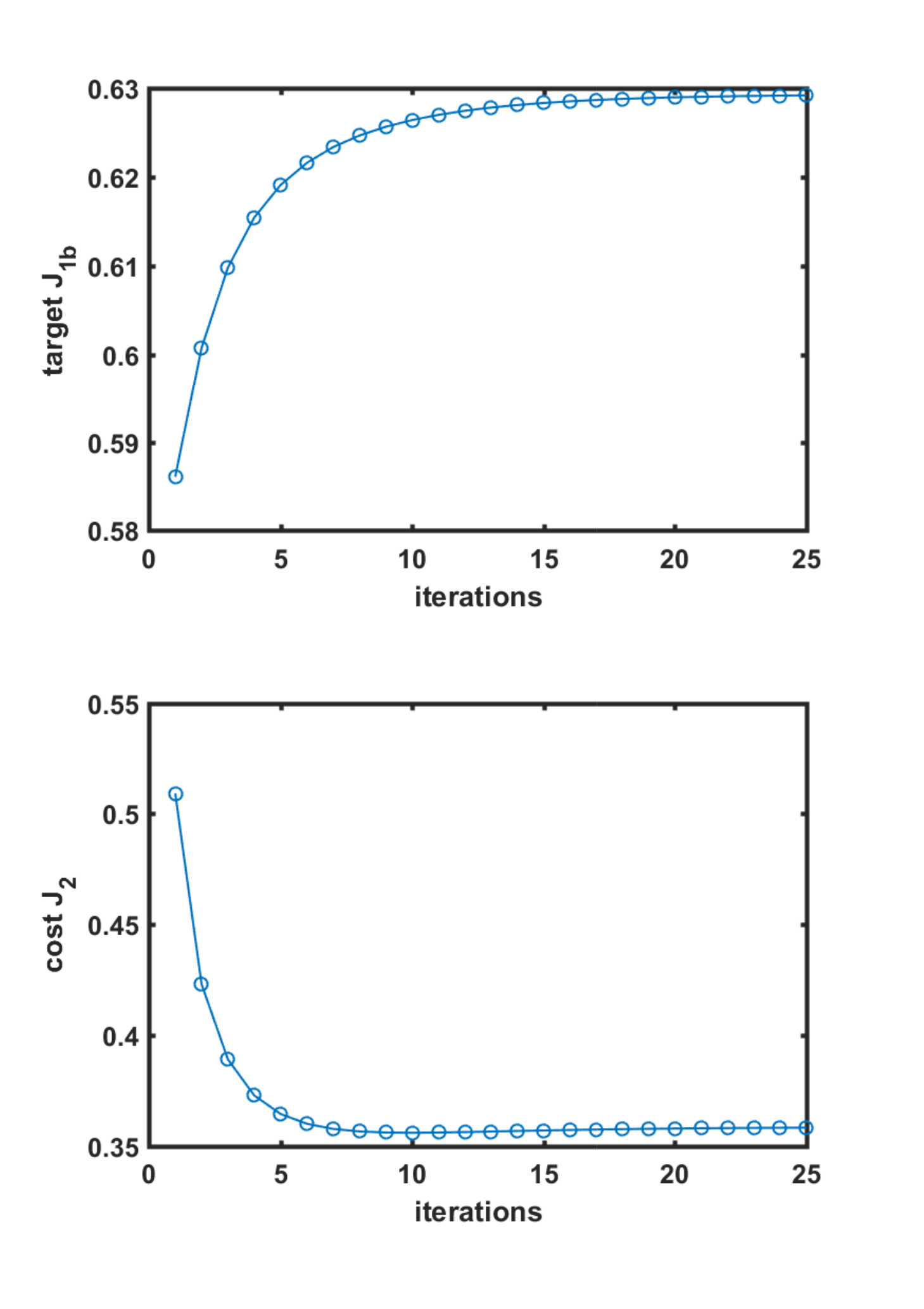}
		\includegraphics[width=5.4cm]{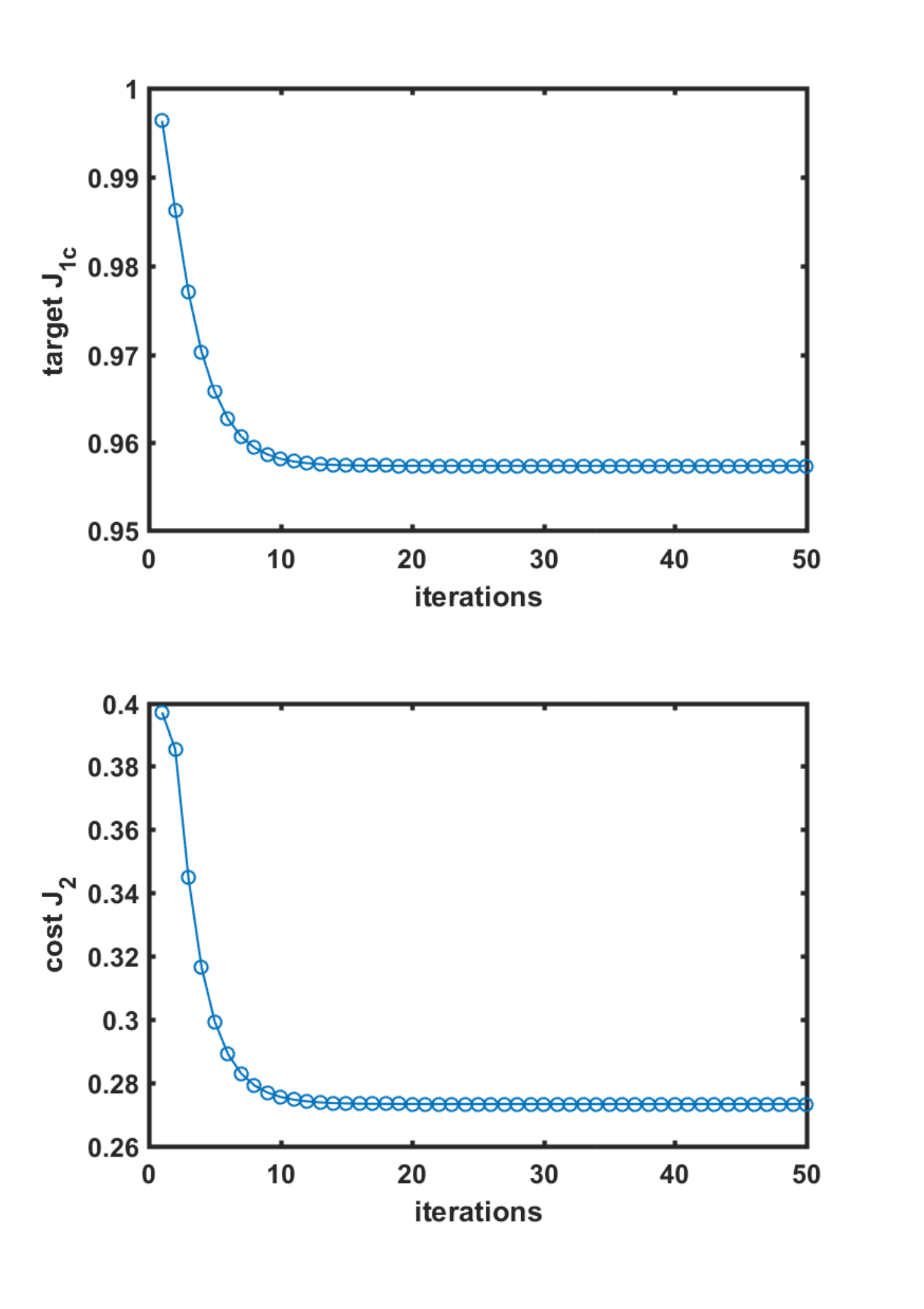}
	\label{fig:qm_optimal}
\end{figure}

Results of a TDSE simulation of a closed quantum system, where the populations are obtained from projection operators using target functional $J_{1a}$ are shown in the left column of Fig.~\ref{fig:qm_optimal}. 
We observe that the population of the target functional (population of $|1\rangle$) rises slowly but continuously during the iterations. 
After 10 cycles it reaches 71 \%, after 100 cycles more than 93 \%. 
At the same time, also the cost functional (fluence of the laser pulse) is rising. 
However, since the former one rises faster than the latter one, the overall control functional is still rising monotonically; for a formal proof, see Ref.~\cite{Zhu1998a}.
The resulting pulse is essentially monochromatic;  due to the prescribed sin$^2$ shape, the envelope has a smooth switch-on and switch-off, and it is approximately symmetric within the 1 ps time window.

The central column  of Fig.~\ref{fig:qm_optimal} shows the same but for an LvNE simulation of an open quantum system, here for $\Gamma_{0\leftarrow 1}=1.24$ ps$^{-1}$.
For the vectorized densities, a target functional of type $J_{1b}$ is optimized.
Even though the cost functional $J_2$ shows that the resulting pulses are more intense, the efficiency of populating the excited state $|1\rangle$ is much lower.
Another major difference lies in the shape of the envelope of the resulting laser pulse. 
Here, the highest amplitude occurs only after approximately three quarters of the prescribed time window of 1 ps. 
Obviously, the optimization avoids creating excited state population too early because it would relax back to the ground state before the final time.

The right column of Fig.~\ref{fig:qm_optimal} shows again a TDSE simulation. 
In contrast to the results shown in the left column of that figure, now the populations are calculated from overlaps using a target functional of type $J_{1c}$, see Eq.~(\ref{eq:target}) and also Sec.~\ref{sec:qm_matrix}.
Already after the first iteration, the population of the target state is already exceeding 99 \%.
During the following iteration steps, the total function $J$ increases further because the cost functional $J_2$ is reduced substantially which is in accordance with results of Ref.~\cite{Zhu1998}.

\section{qm\_balance}
\label{sec:qm_balance}

\subsection{Model reduction and the $\mathcal{H}_2$ error}

The central task in dimension reduction is to find lower-dimensional (reduced) systems which approximate the input-output behavior of a driven dynamical system, see Secs.~\ref{sec:qm_abncd:input} and \ref{sec:qm_abncd:output}, as closely as possible on any compact time interval $[0;T]$. 
In practice this means that the $\mathcal{H}_2$ error norm has to be made as small as possible.
In order to calculate this error, consider the following error system \cite{Benner2012}:
\begin{equation}
A_E=\begin{bmatrix}A&0\\0&\hat{A}\end{bmatrix},\,
N_{k,E}=\begin{bmatrix}N_k&0\\0&\hat{N}_k\end{bmatrix},\,
B_E=\begin{bmatrix}B\\\hat{B}\end{bmatrix},\,
C_E=\begin{bmatrix}C&-\hat{C}\end{bmatrix}
\end{equation}
where the matrices without and with hats stand for the original and the reduced system, respectively. 
Here, as well as in the following Secs.~\ref{sec:qm_truncate} and \ref{sec:qm_H2model}, we will restrict ourselves to the case of open quantum systems modeled in terms of an LvNE, because there model reduction is more important than for TDSE calculations.

Once the error system has been set up, a generalized Lyapunov equation 
\begin{equation}
A_EW_E+W_EA_E^\dagger+\sum_{k=1}^mN_{k,E}W_EN_{k,E}^\dagger+B_EB_E^\dagger=0
\label{eq:glyap_E}
\end{equation}
has to be solved; for remarks how to solve such an equation, see below.
The resulting Gramian $W_E$ can be used to obtain the ${\mathcal{H}_2}$ error norm as follows
\begin{equation}
\mathcal{E}_{\mathcal{H}^2}=\left(C_EW_EC_E^\dagger\right) = \left(B_E^\dagger W_EB_E\right)
\label{eq:error}
\end{equation}
which is often used to quantify the error introduced by dimension reduction.
For example, within \textsc{WavePacket} it can be calculated in the auxiliary  functions \texttt{qm\_H2error} and \texttt{qm\_BTversusH2}. 
Note, however, that the derivation of the $\mathcal{H}_2$ norm is based on output components generated by a $\delta$-like (impulse) input components \cite{Zhou1998}.
Hence, it is often mandatory to also consider the time-dependence generated by specific inputs, see e.~g. our work in Refs.~\cite{BSchmidt:61,BSchmidt:77}.

\subsection{Generalized Lyapunov equations}

The problem of dimension reduction is closely connected to the concepts of controllability and observability (which are dual to each other).
They are characterized in terms of Gramian matrices $W_C$ and $W_O$, respectively. 
Their direct calculation involves a Volterra series with multiply nested time integrals \cite{BSchmidt:61}.
In practice, however, it is of advantage to obtain the Gramians as the symmetric, positive semi-definite solutions of generalized Lyapunov equations. 
For the case of a bilinear input equation (\ref{eq:input:bilinear}) and a linear output equation (\ref{eq:output:linear}) they are given by \cite{Zhang:02a,Bai2006}
\begin{eqnarray}
AW_C+W_C A^\dagger +\sum_{k=1}^mN_kW_CN_k^\dagger+BB^\dagger &=&0 \nonumber \\
A^\dagger W_0+W_0 A+\sum_{k=1}^m N_k^\dagger W_0N_k+C^\dagger C&=&0
\label{eq:glyap_C_O}
\end{eqnarray}
where matrices $B$ and $C$ comprise all vectors $b_k$ and $c_q$, respectively, see Sec.~\ref{sec:qm_abncd}.
Because direct methods for solving such equations have a numerical complexity of $\mathcal{O}(n^6)$, two alternative approaches are implemented in the \textsc{WavePacket} function \texttt{qm\_balance}. 
The first one is based on mapping the Gramian matrices onto vectors.
Then the generalized Lyapunov equations can be understood as systems of coupled linear equations which can be solved, e. g., by means of the bi-conjugate gradient method available in the \texttt{bicg} function of \textsc{Matlab}.
It is advantageous to pre-condition the problem by the solution of the standard Lyapunov equation, i.~e. for $N_k=0$. 
Alternatively, the following iteration can be used \cite{Wachspress1988,Damm2008}
\begin{eqnarray}
AX_0+X_0A^\dagger +BB^\dagger&=&0 \nonumber \\
AX_j+X_jA^\dagger+\sum_{k=1}^m N_kX_{j-1}N^\dagger_k+BB^\dagger&=&0,\,j>0
\label{eq:glyap_iter}
\end{eqnarray}
and similarly for the second (dual) equation for the observability Gramian. 
For $r$ iteration steps this only requires $\mathcal{O}(rn^3)$ operations, because \textsc{WavePacket} solves the standard Lyapunov equations in each step by the
Bartels-Stewart algorithm implemented in function \texttt{lyap} provided with the control toolbox of \textsc{Matlab}. 

Convergence $X_j\rightarrow W_C$ is guaranteed if the eigenvalue of $A$ with the largest (negative) real part is sufficiently separated from the imaginary axis \cite{Wachspress1988}. 
For linear control systems (with $N_k=0$) the Lyapunov equations can always be solved if matrix $A$ is stable, i.~e. having all its eigenvalues in the open left half of the complex number plane.
Stability thus means that there are constants $\lambda, a > 0$ such that $||\exp(At)|| \le \lambda \exp(-at)$.
For systems where this is not the case, e.~g., for the LvNE (\ref{eq:lvne}) with Lindblad dissipation/dephasing, \textsc{WavePacket} offers two numeric stabilization techniques, see App.~\ref{sec:appC}.
For non-linear control systems, where the generalized Lyapunov equations (\ref{eq:glyap_C_O}) have to be solved, the controllability and observability Gramians exist only if
\begin{equation}
\frac{\lambda^2}{2a}\sum_{k=1}^m||N_k||^2<1
\end{equation}
where $||\cdot||$ is the matrix 2-norm induced by the Euclidean norm $|\cdot|$. 
This can be assured by a suitable scaling $u\rightarrow \xi u, N\rightarrow N/\xi, B \rightarrow B/\xi$ with real $\xi>1$ which leaves the equations of motion invariant but not the Gramians. 
Hence, by increasing $\xi$, we can ensure solvability of (\ref{eq:glyap_C_O}).
In our numeric \textsc{WavePacket} experiments reported in Ref.~\cite{BSchmidt:77}, we observe a sensitive dependence on the value of this scaling factor. 
In some cases, good results are obtained only for large scaling factors.
However, because this scaling drives the system toward its linear counterpart, the parameter $\xi$ should not be chosen larger than necessary. 

\subsection{Balancing controllability and observability}

In heuristic approaches to dimension reduction, states are often excluded because they are not reachable by external control fields (not \textit{controllable}) or because they do not contribute to the specified control target (not \textit{observable}).
However, model order reduction can become more challenging when states that are observable are not controllable or vice versa.
This is the motivation for the balancing transformation which strives at finding states which are controllable and observable at the same time.
The idea of such a transformation rests on the transformation properties of a control system, applied to its controllability and observability Gramians \cite{Zhou1998}. 
Upon linear change of coordinates, the state vectors and system matrices transform according to 
\begin{eqnarray}
	\tilde{x}&=&Sx \nonumber \\
	\tilde{A}&=&SAT \nonumber \\
	\tilde{N}_k&=&SN_kT \nonumber \\
	\tilde{b}_k&=&Sb_k \nonumber \\
	\tilde{C}&=&CT
\end{eqnarray}
where T =S$^{-1}$ is the inverse of $S$ (for the case of square, invertible $S$) or the pseudoinverse of $S$ with $STS=S$ and $TST=T$ (if $S$ is singular or rectangular). 
This implies the following transformations for the controllability and observability Gramians
\begin{eqnarray}
\tilde{W}_C&=&SW_CS^\dagger \nonumber \\
\tilde{W}_O&=&T^\dagger W_OT \nonumber \\
\tilde{W}_C\tilde{W}_O&=&SW_CW_OT
\end{eqnarray}
While the eigenvalues of the Gramians $W_C$ and $W_O$ are not invariant, those of the product of the Gramians are invariant under the similarity transformation.

The central idea of balancing is to find a coordinate transformation under which controllability and observability Gramians become equal and diagonal
\begin{equation}
\tilde{W}_C=\tilde{W}_O=\Sigma
\end{equation}
where $\Sigma$ is a diagonal matrix. 
Its elements $\sigma_i>0$ are known as Hankel singular values (HSVs) of the system; they are the square roots of the product of the Gramians. 
Note that the transformation is a contragredient transformation which exists whenever $W_C$ and $W_O$ are symmetric and positive definite \cite{Laub1987}. 
In the balanced representation, states that are least influenced by the input also have the least influence on the output and vice versa, see for an example Fig.~5 of our work in Ref.~\cite{BSchmidt:77}. 
Within  the \textsc{WavePacket} function \texttt{qm\_balance}, there is a choice of two different approaches to find the desired transformation.
One is the "Square Root Balanced Truncation" (SRBT) algorithm \cite{Laub1987,Tombs1987}, the other is the "Minimal Realization and Model Reduction" (MRMR) algorithm \cite{Hahn2002}.

\subsection{Example code}
\label{sec:qm_balance:Example}

Here we are giving a few examples how to set the most important keywords, typically within \texttt{qm\_init}, before running \texttt{qm\_balance} to carry out the balancing transformation.
\begin{verbatim}
reduce.balance.BN_scale = 6; 
reduce.balance.method = 'iter';
reduce.balance.transform = 'srbt'; 
reduce.balance.A_stable = 'ssu';        
reduce.balance.A_split = 1;  
\end{verbatim}
In the first line, the scaling factor $\xi$ for upscaling the field $u(t)$ and downscaling the matrices $B,N$ has been set to 6.
The second line specifies the method of solving the generalized Lyapunov equations (\ref{eq:glyap_C_O}) which can be either '\texttt{iter}' for the iterative solver (\ref{eq:glyap_iter}) or '\texttt{bicg}' for the bi-conjugate gradient method.
In the third line, the SRBT balancing method has been chosen, see above.
Furthermore, the stabilization of the $A$ matrix is achieved by separating the stable from the unstable part ('\texttt{ssu}'), where in the case of LvNE dynamics there is only one unstable component that has to be split off, see also App.~\ref{sec:appC}.
Alternatively, a shift of the eigenvalues ('\texttt{evs}') of the $A$ matrix can be invoked by the following statements 
\begin{verbatim}
reduce.balance.A_stable = 'evs';        
reduce.balance.A_shift = 1e-4; 
\end{verbatim}
with $\alpha=10^{-4}$, see again App.~\ref{sec:appC}.
Finally, it is noted that the WavePacket function \texttt{qm\_balance} reads the original $A,B,N,C$ matrices from an unformatted data file \texttt{lvne.dat} for simulations of open quantum systems, respectively.
Upon transformation, the balanced $\tilde{A},\tilde{B},\tilde{N},\tilde{C}$ matrices are written to a file named \texttt{lvne\_b.mat}, see Fig.~\ref{fig:flowchart}.
\section{qm\_truncate}
\label{sec:qm_truncate}

\subsection{Decomposition of the balanced system}
By its very construction, the balancing transformation implies that those states corresponding to large HSVs ($x_1 \in \mathbf{R}^d$) are more controllable and more observable than those states corresponding to small HSVs ($x_2\in \mathbf{R}^{n-d}$). 
Then the matrix $\Sigma$ can be written as $\Sigma=\left(\Sigma_1,\Sigma_2\right)$ where $\Sigma_1 \in \mathbf{R}^{d\times d}$ and $\Sigma_2 \in \mathbf{R}^{(n-d)\times (n-d)}$, according to the decomposition of the system states into relevant and irrelevant states \cite{hartmann2010e,hartmann2010c}. 
Using corresponding decompositions of $A, b, N, C$, one obtains the coupled equations of motion 
\begin{eqnarray}
\dot{\tilde{x}}_1&=&\tilde{A}_{11}\tilde{x}_1+\tilde{A}_{12}\tilde{x}_2+\sum_{k=1}^m\left(\tilde{N}_{k,11}\tilde{x}_1+\tilde{N}_{k,12}\tilde{x}_2+\tilde{b}_{k,1}\right)u_k \nonumber \\
\dot{\tilde{x}}_2&=&\tilde{A}_{21}\tilde{x}_1+\tilde{A}_{22}\tilde{x}_2+\sum_{k=1}^m\left(\tilde{N}_{k,21}\tilde{x}_1+\tilde{N}_{k,22}\tilde{x}_2+\tilde{b}_{k,2}\right)u_k \nonumber \\
y&=&\tilde{C}_1\tilde{x}_1+\tilde{C}_2\tilde{x}_2
\label{eq:slow_fast}
\end{eqnarray}
In the following, we explain the two approaches implemented in \textsc{WavePacket} function \texttt{qm\_truncate} to deal with the coupled sets of equations.

\subsection{Simple truncation}
The simplest approach to dimension reduction simply consists of a truncation of the less controllable and less observable states $x_2$ 
\begin{eqnarray}
\dot{\tilde{x}}_1&=&\tilde{A}_{11}\tilde{x}_1+\sum_{k=1}^m\left(\tilde{N}_{k,11}\tilde{x}_1+\tilde{b}_{k,1}\right) u_k \nonumber \\
y&=&\tilde{C}_1\tilde{x}_1
\label{eq:truncated}
\end{eqnarray}
This can be justified as being the $\epsilon\rightarrow 0$ limit of (\ref{eq:slow_fast}) assuming $\Sigma_2=\mathcal{O}(\epsilon$) with $0<\epsilon\ll 1$, see Ref.~\cite{BSchmidt:77}.
The truncated subsystem $x_1$ is balanced and stable, and - at least for the case of linear systems - an upper bound for the error of its transfer function is known \cite{Zhou1998}. 

\subsection{Singular perturbation theory}
Alternatively, an averaging principle based on singular perturbation theory can be used, in a similar spirit to the treatment of systems with slow and fast degrees of freedom (dof's) \cite{hartmann2010e,hartmann2010c}. 
Based on the analogy between large HSV-modes with slow dof's and low HSV-modes with fast dof's, we assume that the latter ones to have relaxed to their steady state, $\tilde{x}_2 \rightarrow -\tilde{A}_{22}^{-1}\tilde{A}_{21}\tilde{x}_1$ for the $t \rightarrow \infty$ limit.
Then one can derive equations of motion for $\tilde{x}_1$ which look like (\ref{eq:truncated}) but with the following substitutions
\begin{eqnarray}
\tilde{A}_{11}  &\rightarrow& \tilde{A}_{11}  -\tilde{A}_{12}  \tilde{A}_{22}^{-1}\tilde{A}_{21} \nonumber \\
\tilde{N}_{k,11}&\rightarrow& \tilde{N}_{k,11}-\tilde{N}_{k,12}\tilde{A}_{22}^{-1}\tilde{A}_{21} \nonumber \\
\tilde{C}_{1}   &\rightarrow& \tilde{C}_1     -\tilde{C}_{2}   \tilde{A}_{22}^{-1}\tilde{A}_{21}
\end{eqnarray}
Both the simple truncation and the averaging principle have been implemented in the recent version of the \textsc{WavePacket} function \texttt{qm\_truncate}.
However, in a series of numeric test calculations no clear preference for either one has been found, i.~e. the (moderate) additional effort of the singular perturbation method seems not to lead to more accurate reduced models than the simple truncation \cite{BSchmidt:77}.

\subsection{Example: Asymmetric double well}
\label{sec:qm_truncate:Boris}

Here we will switch to the one-dimensional asymmetric double well system considered also in our previous work \cite{BSchmidt:61,BSchmidt:77}.
In addition to six (five) stationary quantum states which are essentially localized in the left (right) well, we also include the first ten delocalized eigenstates; higher states are neglected for simplicity. 
The 21 considered states lead to a density matrix with dimension $n=441$ thus rendering model order reduction very useful, e.~g., during refinements of fields in quantum optimal control simulations.

The effect of truncation on the time-dependence of observables for given control fields, as well as a comparison of spectra of full versus reduced $A$ matrices has already been shown in Refs.~\cite{BSchmidt:61,BSchmidt:77}. Here we want to present results of quantum optimal control simulations using the \textsc{WavePacket} function \texttt{qm\_optimal} in full versus reduced dimensionality. 
First, we define as observables the total populations of all states in the left (energetically lower) well, in the right well, and of delozalized states above the barrier.
The target of the optimization is the second of these observables, to be reached  after 100 units of time.
The results after 15 optimization cycles can be seen in Fig.~\ref{fig:qm_truncate}, where control fields and observables after optimization are shown. 
For $0 < t < 30$ the field quasi-resonantly drives the system up the ladder of the quantum states localized inside each of the wells which is not reflected by the rather coarse observables.
At later times, the field drives the population to the delocalized states over the barrier.
From there the population finds its way back to the localized states, preferentially those in the right well, probably due to a combination of stimulated emission and dissipation.
Comparing the left and right half of Fig.~\ref{fig:qm_truncate} shows that the optimized field, as well as the induced population dynamics, in full dimensionality ($n=441$) practically coincides with that for a reduced order model ($d=170$), obtained by balanced truncation using the \textsc{WavePacket} functions \texttt{qm\_balance} and \texttt{qm\_truncate}.
Upon further dimension reduction ($d=160$) the optimized field changes qualitatively, thus indicating a limit to dimension reduction for use in optimal control of this model system.

\section{qm\_H2model}
\label{sec:qm_H2model}

\subsection{$\mathcal{H}_2$-optimal model reduction}

This approach to dimension reduction of bilinear control systems is based on the idea of finding an $\mathcal{H}_2$-optimal system that approximates as closely as possible the transfer matrix of the original system, i.~e. minimizing the $\mathcal{H}_2$- error introduced in Eq.~(\ref{eq:error}). 
The method is inspired by the Bilinear Iterative Rational Krylov Algorithm (BIRKA) \cite{Benner2012,BSchmidt:77}. 
Building on the idea of iterative correction, this algorithm aims at fulfilling the first-order necessary optimality conditions, stated in terms of matrix equations. 
This allows to construct the required projection subspaces as solutions to generalized Sylvester equations. 

In practice, the algorithm works as follows. 
The original $n$-dimensional bilinear control system ($A, B, N, C$) is to be approximated by a $d$-dimensional reduced order model ($\hat{A}, \hat{B}, \hat{N}, \hat{C}$). 
Initially chosen by random, the matrices characterizing the reduced system are refined by the following iteration:

\subsection{Generalized Sylvester equations}

In each iteration step, two generalized Sylvester equations have to be solved 
\begin{eqnarray}
AX+X\hat{A}^\dagger +\sum_{k=1}^mN_kX\hat{N}_k^\dagger+B\hat{B}^\dagger&=&0 \nonumber \\
A^\dagger Y+Y\hat{A}+\sum_{k=1}^m N_k^\dagger Y\hat{N}_k-C^\dagger\hat{C}&=&0
\label{eq:gsylv}
\end{eqnarray}
yielding (rectangular) matrices $X,Y \in \mathbf{C}^{n\times d}$.
Note the formal similarity with the generalized Lyapunov equations (\ref{eq:glyap_C_O}) for the calculations of Gramians; however, there is a sign change in the $C^\dagger C$ term. 
Another difference to the generalized Lyapunov equations is that a direct solution of the generalized Sylvester equations requires "only" $\mathcal{O}(d^3n^3)$
operations where $d$ denotes the dimension of the reduced model. 
As for the case of generalized Lyapunov equations, two alternative approaches are available within the \textsc{WavePacket} function \texttt{qm\_H2model}.
One possibility is to rewrite the generalized equations as a linear problem which can be solved, e. g., by the
bi-conjugate gradient method where it is advantageous to use the solutions of the
corresponding ordinary Sylvester equations for pre-conditioning.
Alternatively, iterative methods \cite{Wachspress1988,Damm2008} can be used instead which requires the solution of a standard Sylvester equation in each step
\begin{eqnarray}
AX_0+X_0\hat{A}^\dagger +B\hat{B}^\dagger&=&0 \nonumber \\
AX_j+X_j\hat{A}^\dagger+\sum_{k=1}^m N_kX_{j-1}\hat{N}^\dagger_k+B\hat{B}^\dagger&=&0,\,j>0
\label{eq:gsylv_iter}
\end{eqnarray}
and similarly for the second (dual) equation for $Y$. 
Note that the cost for the solution of a standard Sylvester equations is less than for a standard Lyapunov equations also because the former ones can be handled efficiently for sparse system matrices. 
Hence, a single step of the BIRKA iteration is computationally less expensive than performing a balancing  transformation.
However, the effort for BIRKA obviously depends on the number of iteration steps required until the fixed point iteration is (numerically) converged. 
Based on the numerical examples studied in Ref.~\cite{BSchmidt:77}, we can not report significant differences
between the methods.
    
\subsection{Fixed point iteration}

Once the generalized Sylvester equations have been solved, a QR-decomposition of matrices $X,Y$ is performed
\begin{eqnarray} 
X&=&VR \nonumber \\
Y&=&WZ
\end{eqnarray}
where $V,W$ are orthogonal matrices and $R,Z$ are upper triangular matrices (not needed here). 
Then $V,W$ are used to construct a refined system in the following way
\begin{eqnarray}
\hat{A}&=&SAV \nonumber \\
\hat{N}_k&=&SN_kV \nonumber \\
\hat{B}&=&SB \nonumber \\
\hat{C}&=&CV
\label{eq:refined}
\end{eqnarray}
which corresponds to a Petrov-Galerkin projection of the original model with
\begin{equation}
	S=\left(W^\dagger V\right)^{-1}W^\dagger
\end{equation}
Then the refined system (\ref{eq:refined}) is inserted into the generalized Sylvester equations (\ref{eq:gsylv}) yielding new matrices $V,W$.
This iteration is repeated until the change in the spectrum of the reduced order system matrix $\hat{A}$ falls below a user prescribed tolerance.

\subsection {Example code}
\label{sec:qm_H2model:Example}
Here we give examples for keywords to be used with the function \texttt{qm\_H2model} for $\mathcal{H}_2$-optimal  model reduction
\begin{verbatim}
reduce.H2model.A_stable = 'ssu';
reduce.H2model.A_split = 1;
reduce.H2model.BN_scale = 3;
reduce.H2model.method = 'iter';
reduce.H2model.max_iter = 200; 
reduce.H2model.conv_tol = 1e-6;
\end{verbatim}
The first three lines concerning the stabilization of $A$-matrix (by splitting off one unstable component) and the scaling of $u,B,N$ are in close analogy to the keywords used for the balancing transformation, see Sec.~\ref{sec:qm_balance:Example}.
The fourth line is to indicate the use of the iterative solver (\ref{eq:gsylv_iter}) for generalized Sylvester equations (\ref{eq:gsylv}).
The last two lines specify termination criteria for the BIRKA iteration process, i.~e., the maximum number of iterations and a convergence tolerance.
Finally, it is noted that the WavePacket function \texttt{qm\_H2model} reads the original $A,B,N,C$ matrices from an unformatted data file \texttt{lvne.dat} for simulations open quantum systems, respectively.
Upon transformation, the reduced $\hat{A},\hat{B},\hat{N},\hat{C}$ matrices are written to a file named \texttt{lvne\_h.mat}, see also Fig.~\ref{fig:flowchart}.

For numeric experiments concerning the accuracy and the computation effort of the $\mathcal{H}_2$-optimal  model reduction, we refer  the reader to Refs.~\cite{Benner2012,BSchmidt:77}.
We have also repeated the calculations presented in Sec.~\ref{sec:qm_truncate:Boris} for the asymmetric double well model.
This time we have compared optimized fields obtained for driven population dynamics for the full-dimensional ($n=441$) model versus an $\mathcal{H}_2$-optimal model, with very similar results: 
We found excellent agreement for $d=170$, but notable deviations for $d=160$, see also Fig.~\ref{fig:qm_truncate}.

\section{Summary and Outlook}
\label{sec:summary}

Complementary to Part~I which described the PDE based solvers for Schr\"{o}dinger equations together with the underlying discrete variable representations implemented in \textsc{WavePacket}, the present Part~II discusses the ODE based approach to quantum dynamics in that program package, which allows to treat Schr\"{o}dinger and Liouville-von Neumann equations on an equal footing.
While so far only Lindblad-Kossakowski models for dissipation and dephasing have been implemented, also Redfield or other approaches would be straight-forward to be include.
This generalization of the \textsc{WavePacket} codes has also allowed for a relatively easy implementation of the rapid monotonically convergent algorithms for optimal control of closed and open quantum systems.
It can be expected that more recent developments in this field, such as multi-target optimization, time-averaged targets and also non-linear models for the interaction between quantum systems and external fields will become available in future versions of our software package.

While most of the above aspects are also available in other packages such as \textsc{MCTDH}, \textsc{TDDVR}, \textsc{QuTiP} and/or \textsc{QLib}, an advantage of \textsc{WavePacket} is that it offers a coherent combination of all these features, sharing the \textsc{Matlab} functions for features such as the definition of the Hamiltonian etc. Moreover, it is emphasized that the target systems for \textsc{WavePacket} are low- to medium-dimensional (model) systems where computational requirements are not the dominant concern, but user-friendliness and on-the-fly graphics are. For example, the FFT-based representations typically allow for $3\ldots 5$ dimensional propagations using \texttt{qm\_propa} or $2\ldots 3$ dimensional bound state calculations using \texttt{qm\_bound} on a standard PC equipped with 8 GB of memory, see also Sec.~7 of Part~I. Alternatively, for dynamical simulations in energy representation using \texttt{qm\_control}, the present version of \textsc{WavePacket} can typically handle thousands of quantum states.
Even though these limits can be pushed forward -- to some extent -- by resorting to more powerful hardware,   the present software certainly does not aim at competing with the MCTDH or TDDVR package with respect to high dimensionality. However, we stress the role of the model order reduction available since version 5.0 of \textsc{WavePacket}.
Three different approaches, i.~e., balanced truncation, singular perturbation theory, and $\mathcal{H}_2$-optimal model reduction, have been implemented.
So far, none of them has been shown to be superior to the others in our numeric experiments, despite of the quite different nature of the approaches.
It remains to be seen how these algorithms perform for different classes of quantum systems.
We hope that the public release will help to spread these algorithms in the research community, and we are expecting feedback from the users which will be also instrumental for the further development.

Since the advent of versions 4.x and 5.x, the \textsc{WavePacket} framework is developed in \textsc{Matlab}.
Despite of limitations in the availability in some academic institutions, we chose that programming environment because it offers several unique features.
There are built-in functions for many frequently used tasks, in particular in the field of numeric linear algebra, including support for sparse matrices, thus allowing for fast code development.
Note that \textsc{Matlab} is rather intuitive to use, due to the close proximity between physical/mathematical formulations and the program codes.
Furthermore, it offers an easy extension of core functionality through function handles, thus making it easy to apply the \textsc{MATLAB} version of \textsc{WavePacket} to different physical situations.
Moreover, graphical output, partly in the form of animations, is readily available to the user, which is helpful to develop a more intuitive understanding quantum dynamics and quantum optimal control.

In this context it may also be of interest how the choice of \textsc{Matlab} affects the performance of \textsc{WavePacket}. To this end, we compared the MATLAB version (described in the present work)  with the C++ version (currently under development).   
For both versions most of the computational time is spent in external libraries such as FFTW, BLAS, so that the result strongly depends on the respective implementation of these libraries. Despite of the obvious limitations, we did a crude comparison between the Matlab and the C++ version. Contrary to common belief, our preliminary results show that the performance of the two versions is similar. Typically simulations with the \textsc{Matlab} version of \texttt{qm\_propa} using an FFT-based DVR/FBR schemes with 32 points in each  dimensions take 41 (3D), 244 (4D), or 10600 (5D) seconds, for 2000 steps (Strang splitting) on a PC with Intel Xeon CPU (E3-1241 v3 @ 3.5 GHz). Finally, it is mentioned that due to the extensive use of advanced \textsc{Matlab} features (cell arrays, structures, occasionally also classes), the \textsc{WavePacket} software does not operate correctly under Gnu Octave.

For the next main version of \textsc{WavePacket} it is planned to go beyond purely quantum-mechanical propagations by also offering functions for classical \cite{BSchmidt:38} and  mixed quantum-classical dynamics \cite{BSchmidt:39,BSchmidt:43}, including surface hopping algorithms \cite{BSchmidt:39,Subotnik2016}. 
Since these generalizations are difficult to be implemented in a completely procedural way, the further development of the \textsc{Matlab} version of \textsc{WavePacket} will be directed toward object-oriented approaches.
Note that first steps have already been implemented for the realizations of the DVR/FBR techniques described in Part~I.
At the same time, the above-mentioned C++ version aims at a rewrite of the \textsc{WavePacket} codes in a completely object-oriented manner.
However, as long as that project is still in an early stage, the \textsc{Matlab} codes presented here will remain the main working version of \textsc{WavePacket} for the next few years. 

Since 2008, the development of the free and open-source \textsc{WavePacket} is hosted at \textsc{SourceForge}, with the version described in this work being 5.3.0.
In addition to an SVN repository providing a central location to manage the distributed development of our software package, there is also a large number of Wiki pages containing complete descriptions of all the \textsc{Matlab} functions, classes, and variables, serving as a reference to users.
Further information about the physical and numerical background is also available on the Wiki pages of the \textsc{WavePacket} main project, along with a large number of worked out demonstration examples, complete with input and output data files, often including animated graphics as MP4 files.
These examples can also be understood as a tutorial, not only including all examples presented in Parts~I and II of this work, but also demonstrating the use of our software package for model systems beyond one dimension, both for single and coupled Schr\"{o}dinger equations.

\begin{acknowledgments}
This work has been supported by the Einstein Center for Mathematics Berlin (ECMath) through projects SE~11 and SE~20.
Boris Sch\"afer-Bung (formerly at FU Berlin) and Tobias Breiten (U of Graz, Austria) are acknowledged for setting up initial versions of the \textsc{MATLAB} codes for dimension reduction.
Finally, we are grateful to Ulf Lorenz (formerly at U Potsdam) for his valuable help with all kind of questions around the \textsc{WavePacket} software package.
\end{acknowledgments}

\begin{appendix}
\section{Compatibility issues}
\label{sec:appA}

With the introduction of versions 5.3.0 a few minor backward incompatibilities have arisen.
In particular there are a few changes in the initialization (normally provided through self-written Matlab function \texttt{qm\_init.m}) with respect to version 5.2.3 described in Part~I.
These changes have become necessary because we decided to have (almost) all variables as \textsc{Matlab} structures with three hierarchical  levels, see Tab.~\ref{tab:fieldnames}.
Moreover, the class definitions for DVR/FBR grids and (associated) kinetic operators have been arranged into package folders, in order to be organized similarly to potentials, dipoles, etc. Hence, in the respective notations, underscores have to be replaced by dots, e.~g. \texttt{grid\_fft} has to be replaced by \texttt{grid.fft}. 
For more details about these syntax changes, the reader is referred to the news section in the Wiki pages of the \textsc{Matlab} version of \textsc{WavePacket}.

\begin{table}
\begin{tabular}{ll}
\hline \hline
 until V5.2.3 & since V5.3.0 \\
\hline
 hamilt.pot.params.xyz & hamilt.pot.xyz \\
 hamilt.dip.params.xyz & hamilt.dip.xyz \\
 hamilt.nip.params.xyz & hamilt.nip.xyz \\
 time.propa.params.xyz & time.propa.xyz \\
 psi.init.dof\{1\}.xyz & psi.dof\{1\}.xyz \\
\hline \hline
\end{tabular}
\caption{Change of fieldnames between different WavePacket versions where \texttt{xyz} stands for various possible field names}
\label{tab:fieldnames}
\end{table}

\section{Dissipation and dephasing models}
\label{sec:appB}

Within the \textsc{WavePacket} software package, population relaxation (dissipation) and associated dephasing are described by Eq.~(\ref{eq:lvne}) using Lindblad operators 
\begin{equation}
\hat L_\ell=\hat L_{ij}=\sqrt{\Gamma_{i\leftarrow j}}|i\rangle\langle j|
\label{eq:projector}
\end{equation}
where the summation extends over all possible channels $\ell=(i\leftarrow j)$. 
With the (phenomenological) rate constants given as inverse times $\Gamma_{i\leftarrow j}=1/T_{i\leftarrow j}$ the Lindblad evolution is trace-preserving, i.e. the sum of populations remains constant, and completely positive, i. e., also the individual populations remain positive.
Typically, the upward rates are calculated from the downward ones using the principle of detailed balance
\begin{equation}
\Gamma_{j\leftarrow i}=\exp\left( -\frac{E_j-E_i}{k_B\Theta}\right )\Gamma_{i\leftarrow j}, \quad j>i
\label{eq:balance}
\end{equation}
which ensures that the densities approach the Boltzmann distribution for temperature $\Theta$ in the limit of infinitely long times.

Specific models for the rate constants defined in Eq.~(\ref{eq:projector}) require - in principle - a microscopic knowledge of the system-bath coupling operator. 
Since this information is usually not available, simplifying assumptions have to be made.
Two such models are currently available within the \texttt{WavePacket} function \texttt{qm\_abncd}.
The first one builds on the assumption that the system--bath coupling Hamiltonian is linear in the bath modes, see Eq.~(\ref{eq:sbc}).
Using Fermi's golden rule for the weak coupling limit, and assuming equal masses and frequencies of the bath modes, it can be shown that the downward (population) relaxation rates fulfill the following relation (with $\omega_{ji} \equiv E_j-E_i$)
\begin{equation}
\label{eq:fermi}
\Gamma_{i\leftarrow j}\propto \frac{\chi_{ji}^2}{\omega_{ji}}\frac{1}{\exp{\frac{\omega_{ji}}{k_BT}}-1},\quad  j>i
\end{equation}
see e.~g. Ref.~\cite{Andrianov2006} for an application to molecular vibrations.
Another frequently used model is based on scaled Einstein coefficients for spontaneous emission 
\begin{equation}
\label{eq:einstein}
\Gamma_{i\leftarrow j}\propto\mu_{ji}^2\omega_{ji}^3 ,\quad  j>i
\end{equation}
see Ref.~\cite{Tremblay2011a} for an application to molecular electronic dynamics.
Together with the principle of detailed balance (\ref{eq:balance}), these two models only require the specification of one relaxation rate (typically $\Gamma_{0\leftarrow 1}$) to determine all rates of a Lindblad model (\ref{eq:lvne}).
As an alternative to these scaling relations, \textsc{WavePacket} also offers the possibility to assume constant relaxation rates or to read pre-computed rates from input data files. 

Within the \textsc{WavePacket} software package, a frequently used model for pure dephasing can be described within the Lindblad model of Eq.~(\ref{eq:lvne}) with operators
\begin{equation}
\hat{L}_\ell=\sqrt{2\kappa}\sum_\ell|\ell\rangle E_\ell\langle \ell|=\sqrt{2\kappa}\hat{H}_0
\end{equation}
where $\kappa>0$ is a scaling factor.
This leads to a quadratic energy gap dependence for the dephasing rate
\begin{equation}
\gamma_{ij}^\ast=\kappa\omega_{ij}^2
\end{equation}
see, e.~g., Ref.~\cite{Lockwood2001} for an application on vibrational dephasing rates for molecules interacting with a bath.
Alternatively, \textsc{WavePacket} also offers the possibility to assume constant dephasing rates or to read those rates from input data files. 

\section{Stabilizing the $A$ matrix}
\label{sec:appC}

In stability theory, a stable system approaches a fixed point (an equilibrium) in the long time limit, and nearby points converge to it at an exponential rate. In the input equations (\ref{eq:input:linear},\ref{eq:input:bilinear}) given in Sec.~\ref{sec:qm_abncd:input}, assuming vanishing fields $u(t)=0$, this requires that the spectrum of the system matrix $A$ should be in the left half of the complex number plane (negative real part). 
Such matrices are also referred to as Hurwitz stable matrices. 

However, for open quantum system dynamics we use the LvNE (\ref{eq:lvne}) with Lindblad superoperators describing relaxation to the thermal equilibrium $x_e$. 
Hence, matrix $A$ has a simple eigenvalue zero.
In such cases, the $A$ matrix can be stabilized by one of the following two techniques:
\begin{itemize}
\item The diagonal values of matrix A can be shifted by a small negative amount, $A \rightarrow A-\alpha I$, where $\alpha>0$ is a real-valued shift parameter. 
Solutions of ODEs always contain a term of the form $\exp(At)$, hence this shift introduces a damping of the form $\exp(-\alpha t)$. 
In optimal control theory this is referred to as "discounting a functional", i.e., the further future is not taken quite as important as the closer future. 
This damping drives the system towards $x=0$, i.e., to the equilibrium state, even for  the case of closed quantum system dynamics described by TDSE, where this procedure may violate norm conservation.
\item The unstable part of $A$ can be separated by transforming the matrices $A, N_k, C$ and the vectors $b_k, x(t)$ into the eigenbasis of $A$. If we order the eigenvalues by the absolute value of their real parts, we can directly separate the unstable part $x_1\in C^M$ from the stable part $x_2 \in C^{n \times (n-M)}$. 
Since a straightforward implementation will -- in general -- destroy the sparsity pattern of the matrices involved,
we suggest to use a particular technique for sparsity preserving projections, see Chap.~5.2 of Ref.~\cite{BSchmidt:77}.
Note that in the case of open quantum system dynamics described by LvNE  (\ref{eq:lvne}) with Lindblad superoperators it is sufficient to choose $M=1$, i.e. there is only one unstable component (eigenvalue zero) to be removed.
\end{itemize}

\end{appendix}

\bibliography{WavePacket2}

\begin{thebibliography}{68}
\expandafter\ifx\csname natexlab\endcsname\relax\def\natexlab#1{#1}\fi
\expandafter\ifx\csname bibnamefont\endcsname\relax
  \def\bibnamefont#1{#1}\fi
\expandafter\ifx\csname bibfnamefont\endcsname\relax
  \def\bibfnamefont#1{#1}\fi
\expandafter\ifx\csname citenamefont\endcsname\relax
  \def\citenamefont#1{#1}\fi
\expandafter\ifx\csname url\endcsname\relax
  \def\url#1{\texttt{#1}}\fi
\expandafter\ifx\csname urlprefix\endcsname\relax\def\urlprefix{URL }\fi
\providecommand{\bibinfo}[2]{#2}
\providecommand{\eprint}[2][]{\url{#2}}

\bibitem[{\citenamefont{May and K{\"{u}}hn}(2000)}]{May:00a}
\bibinfo{author}{\bibfnamefont{V.}~\bibnamefont{May}} \bibnamefont{and}
  \bibinfo{author}{\bibfnamefont{O.}~\bibnamefont{K{\"{u}}hn}},
  \emph{\bibinfo{title}{{Charge and Energy Transfer Dynamics in Molecular
  Systems}}} (\bibinfo{publisher}{Wiley}, \bibinfo{address}{Berlin},
  \bibinfo{year}{2000}).

\bibitem[{\citenamefont{Schleich}(2001)}]{Schleich:01a}
\bibinfo{author}{\bibfnamefont{W.~P.} \bibnamefont{Schleich}},
  \emph{\bibinfo{title}{{Quantum Optics in Phase Space}}}
  (\bibinfo{publisher}{Wiley--VCH}, \bibinfo{address}{Berlin},
  \bibinfo{year}{2001}).

\bibitem[{\citenamefont{Tannor}(2004)}]{Tannor:04a}
\bibinfo{author}{\bibfnamefont{D.}~\bibnamefont{Tannor}},
  \emph{\bibinfo{title}{{Introduction to Quantum Mechanics. A Time-Dependent
  Perspective}}} (\bibinfo{publisher}{University Science Books},
  \bibinfo{address}{Sausalito}, \bibinfo{year}{2004}).

\bibitem[{\citenamefont{Gro{\ss}mann}(2008)}]{Grossmann2008}
\bibinfo{author}{\bibfnamefont{F.}~\bibnamefont{Gro{\ss}mann}},
  \emph{\bibinfo{title}{{Theoretical Femtosecond Physics}}}
  (\bibinfo{publisher}{Springer}, \bibinfo{address}{Berlin Heidelberg},
  \bibinfo{year}{2008}).

\bibitem[{\citenamefont{Zewail}(2000)}]{Zewail2000}
\bibinfo{author}{\bibfnamefont{A.~H.} \bibnamefont{Zewail}},
  \bibinfo{journal}{The Journal of Physical Chemistry A}
  \textbf{\bibinfo{volume}{104}}, \bibinfo{pages}{5660} (\bibinfo{year}{2000}).

\bibitem[{\citenamefont{Sundstr{\"{o}}m}(2008)}]{Sundstrom2008}
\bibinfo{author}{\bibfnamefont{V.}~\bibnamefont{Sundstr{\"{o}}m}},
  \bibinfo{journal}{Annual Review of Physical Chemistry}
  \textbf{\bibinfo{volume}{59}}, \bibinfo{pages}{53} (\bibinfo{year}{2008}).

\bibitem[{\citenamefont{de~Vivie-Riedle and Troppmann}(2007)}]{Vivie:07a}
\bibinfo{author}{\bibfnamefont{R.}~\bibnamefont{de~Vivie-Riedle}}
  \bibnamefont{and}
  \bibinfo{author}{\bibfnamefont{U.}~\bibnamefont{Troppmann}},
  \bibinfo{journal}{Chemical Reviews} \textbf{\bibinfo{volume}{107}},
  \bibinfo{pages}{5082} (\bibinfo{year}{2007}).

\bibitem[{\citenamefont{Zhu et~al.}(2013)\citenamefont{Zhu, Kais, Wei,
  Herschbach, and Friedrich}}]{Zhu2013}
\bibinfo{author}{\bibfnamefont{J.}~\bibnamefont{Zhu}},
  \bibinfo{author}{\bibfnamefont{S.}~\bibnamefont{Kais}},
  \bibinfo{author}{\bibfnamefont{Q.}~\bibnamefont{Wei}},
  \bibinfo{author}{\bibfnamefont{D.}~\bibnamefont{Herschbach}},
  \bibnamefont{and}
  \bibinfo{author}{\bibfnamefont{B.}~\bibnamefont{Friedrich}},
  \bibinfo{journal}{The Journal of Chemical Physics}
  \textbf{\bibinfo{volume}{138}}, \bibinfo{pages}{024104}
  (\bibinfo{year}{2013}).

\bibitem[{\citenamefont{Kais}(2014)}]{Kais2014}
\bibinfo{editor}{\bibfnamefont{S.}~\bibnamefont{Kais}}, ed.,
  \emph{\bibinfo{title}{{Quantum Information and Computation for Chemistry}}}
  (\bibinfo{year}{2014}).

\bibitem[{\citenamefont{Glaser et~al.}(2015)\citenamefont{Glaser, Boscain,
  Calarco, Koch, K{\"{o}}ckenberger, Kosloff, Kuprov, Luy, Schirmer,
  Schulte-Herbr{\"{u}}ggen et~al.}}]{Glaser2015}
\bibinfo{author}{\bibfnamefont{S.~J.} \bibnamefont{Glaser}},
  \bibinfo{author}{\bibfnamefont{U.}~\bibnamefont{Boscain}},
  \bibinfo{author}{\bibfnamefont{T.}~\bibnamefont{Calarco}},
  \bibinfo{author}{\bibfnamefont{C.~P.} \bibnamefont{Koch}},
  \bibinfo{author}{\bibfnamefont{W.}~\bibnamefont{K{\"{o}}ckenberger}},
  \bibinfo{author}{\bibfnamefont{R.}~\bibnamefont{Kosloff}},
  \bibinfo{author}{\bibfnamefont{I.}~\bibnamefont{Kuprov}},
  \bibinfo{author}{\bibfnamefont{B.}~\bibnamefont{Luy}},
  \bibinfo{author}{\bibfnamefont{S.}~\bibnamefont{Schirmer}},
  \bibinfo{author}{\bibfnamefont{T.}~\bibnamefont{Schulte-Herbr{\"{u}}ggen}},
  \bibnamefont{et~al.}, \bibinfo{journal}{The European Physical Journal D}
  \textbf{\bibinfo{volume}{69}}, \bibinfo{pages}{279} (\bibinfo{year}{2015}).

\bibitem[{\citenamefont{Beck et~al.}(2000)\citenamefont{Beck, J{\"{a}}ckle,
  Worth, and Meyer}}]{Beck2000}
\bibinfo{author}{\bibfnamefont{M.~H.} \bibnamefont{Beck}},
  \bibinfo{author}{\bibfnamefont{A.}~\bibnamefont{J{\"{a}}ckle}},
  \bibinfo{author}{\bibfnamefont{G.~A.} \bibnamefont{Worth}}, \bibnamefont{and}
  \bibinfo{author}{\bibfnamefont{H.-D.} \bibnamefont{Meyer}},
  \bibinfo{journal}{Physics Reports} \textbf{\bibinfo{volume}{324}},
  \bibinfo{pages}{1} (\bibinfo{year}{2000}).

\bibitem[{\citenamefont{Khan et~al.}(2014)\citenamefont{Khan, Sardar, Sarkar,
  and Adhikari}}]{Khan2014}
\bibinfo{author}{\bibfnamefont{B.~A.} \bibnamefont{Khan}},
  \bibinfo{author}{\bibfnamefont{S.}~\bibnamefont{Sardar}},
  \bibinfo{author}{\bibfnamefont{P.}~\bibnamefont{Sarkar}}, \bibnamefont{and}
  \bibinfo{author}{\bibfnamefont{S.}~\bibnamefont{Adhikari}},
  \bibinfo{journal}{The Journal of Physical Chemistry A}
  \textbf{\bibinfo{volume}{118}}, \bibinfo{pages}{11451}
  (\bibinfo{year}{2014}).

\bibitem[{\citenamefont{Johansson et~al.}(2012)\citenamefont{Johansson, Nation,
  and Nori}}]{Johansson2012}
\bibinfo{author}{\bibfnamefont{J.~R.} \bibnamefont{Johansson}},
  \bibinfo{author}{\bibfnamefont{P.~D.} \bibnamefont{Nation}},
  \bibnamefont{and} \bibinfo{author}{\bibfnamefont{F.}~\bibnamefont{Nori}},
  \bibinfo{journal}{Computer Physics Communications}
  \textbf{\bibinfo{volume}{183}}, \bibinfo{pages}{1760} (\bibinfo{year}{2012}).

\bibitem[{\citenamefont{Johansson et~al.}(2013)\citenamefont{Johansson, Nation,
  and Nori}}]{Johansson2013}
\bibinfo{author}{\bibfnamefont{J.~R.} \bibnamefont{Johansson}},
  \bibinfo{author}{\bibfnamefont{P.~D.} \bibnamefont{Nation}},
  \bibnamefont{and} \bibinfo{author}{\bibfnamefont{F.}~\bibnamefont{Nori}},
  \bibinfo{journal}{Computer Physics Communications}
  \textbf{\bibinfo{volume}{184}}, \bibinfo{pages}{1234} (\bibinfo{year}{2013}).

\bibitem[{\citenamefont{Mendl}(2011)}]{Mendl2011}
\bibinfo{author}{\bibfnamefont{C.~B.} \bibnamefont{Mendl}},
  \bibinfo{journal}{Computer Physics Communications}
  \textbf{\bibinfo{volume}{182}}, \bibinfo{pages}{1327} (\bibinfo{year}{2011}).

\bibitem[{\citenamefont{Machnes et~al.}(2011)\citenamefont{Machnes, Sander,
  Glaser, Fouqui, Gruslys, and Schirmer}}]{Machnes2011}
\bibinfo{author}{\bibfnamefont{S.}~\bibnamefont{Machnes}},
  \bibinfo{author}{\bibfnamefont{U.}~\bibnamefont{Sander}},
  \bibinfo{author}{\bibfnamefont{S.~J.} \bibnamefont{Glaser}},
  \bibinfo{author}{\bibfnamefont{P.~D.} \bibnamefont{Fouqui}},
  \bibinfo{author}{\bibfnamefont{A.}~\bibnamefont{Gruslys}}, \bibnamefont{and}
  \bibinfo{author}{\bibfnamefont{S.}~\bibnamefont{Schirmer}},
  \bibinfo{journal}{Physical Review A} \textbf{\bibinfo{volume}{84}},
  \bibinfo{pages}{022305} (\bibinfo{year}{2011}).

\bibitem[{\citenamefont{Schmidt and Lorenz}(2017)}]{BSchmidt:75}
\bibinfo{author}{\bibfnamefont{B.}~\bibnamefont{Schmidt}} \bibnamefont{and}
  \bibinfo{author}{\bibfnamefont{U.}~\bibnamefont{Lorenz}},
  \bibinfo{journal}{Computer Physics Communications}
  \textbf{\bibinfo{volume}{213}}, \bibinfo{pages}{223} (\bibinfo{year}{2017}).

\bibitem[{\citenamefont{Light et~al.}(1985)\citenamefont{Light, Hamilton, and
  Lill}}]{Light:85a}
\bibinfo{author}{\bibfnamefont{J.~C.} \bibnamefont{Light}},
  \bibinfo{author}{\bibfnamefont{I.~P.} \bibnamefont{Hamilton}},
  \bibnamefont{and} \bibinfo{author}{\bibfnamefont{J.~V.} \bibnamefont{Lill}},
  \bibinfo{journal}{The Journal of Chemical Physics}
  \textbf{\bibinfo{volume}{82}}, \bibinfo{pages}{1400} (\bibinfo{year}{1985}).

\bibitem[{\citenamefont{Light and Carrington}(2000)}]{Light:00a}
\bibinfo{author}{\bibfnamefont{J.~C.} \bibnamefont{Light}} \bibnamefont{and}
  \bibinfo{author}{\bibfnamefont{T.}~\bibnamefont{Carrington}},
  \bibinfo{journal}{Advances in Chemical Physics}
  \textbf{\bibinfo{volume}{114}}, \bibinfo{pages}{263} (\bibinfo{year}{2000}).

\bibitem[{\citenamefont{Leforestier et~al.}(1991)\citenamefont{Leforestier,
  Bisseling, Cerjan, Feit, Friesner, Guldberg, Hammerich, Jolicard, Karrlein,
  Meyer et~al.}}]{Leforestier:91a}
\bibinfo{author}{\bibfnamefont{C.}~\bibnamefont{Leforestier}},
  \bibinfo{author}{\bibfnamefont{R.}~\bibnamefont{Bisseling}},
  \bibinfo{author}{\bibfnamefont{C.}~\bibnamefont{Cerjan}},
  \bibinfo{author}{\bibfnamefont{M.}~\bibnamefont{Feit}},
  \bibinfo{author}{\bibfnamefont{R.}~\bibnamefont{Friesner}},
  \bibinfo{author}{\bibfnamefont{A.}~\bibnamefont{Guldberg}},
  \bibinfo{author}{\bibfnamefont{A.}~\bibnamefont{Hammerich}},
  \bibinfo{author}{\bibfnamefont{G.}~\bibnamefont{Jolicard}},
  \bibinfo{author}{\bibfnamefont{W.}~\bibnamefont{Karrlein}},
  \bibinfo{author}{\bibfnamefont{H.-D.} \bibnamefont{Meyer}},
  \bibnamefont{et~al.}, \bibinfo{journal}{Journal of Computational Physics}
  \textbf{\bibinfo{volume}{94}}, \bibinfo{pages}{59} (\bibinfo{year}{1991}).

\bibitem[{\citenamefont{Baer}(2006)}]{Baer:06a}
\bibinfo{author}{\bibfnamefont{M.}~\bibnamefont{Baer}},
  \emph{\bibinfo{title}{{Beyond Born--Oppenheimer}}}
  (\bibinfo{publisher}{Wiley-VCH}, \bibinfo{address}{Hoboken, New Jersey},
  \bibinfo{year}{2006}).

\bibitem[{\citenamefont{Domcke et~al.}(2004)\citenamefont{Domcke, Yarkony, and
  K{\"{o}}ppel}}]{Domcke:04a}
\bibinfo{editor}{\bibfnamefont{W.}~\bibnamefont{Domcke}},
  \bibinfo{editor}{\bibfnamefont{D.~R.} \bibnamefont{Yarkony}},
  \bibnamefont{and}
  \bibinfo{editor}{\bibfnamefont{H.}~\bibnamefont{K{\"{o}}ppel}}, eds.,
  \emph{\bibinfo{title}{{Conical Intersections. Electronic Structure, Dynamics
  and Spectroscopy}}}, vol.~\bibinfo{volume}{15} of
  \emph{\bibinfo{series}{Advanced Series in Physical Chemistry}}
  (\bibinfo{publisher}{World Scientific}, \bibinfo{address}{Singapore},
  \bibinfo{year}{2004}).

\bibitem[{\citenamefont{Weiss}(1999)}]{Weiss:99a}
\bibinfo{author}{\bibfnamefont{U.}~\bibnamefont{Weiss}},
  \emph{\bibinfo{title}{{Quantum Dissipative Systems}}}
  (\bibinfo{publisher}{World Scientific}, \bibinfo{address}{Singapore},
  \bibinfo{year}{1999}).

\bibitem[{\citenamefont{Breuer and Petruccione}(2002)}]{Breuer:02a}
\bibinfo{author}{\bibfnamefont{H.-P.} \bibnamefont{Breuer}} \bibnamefont{and}
  \bibinfo{author}{\bibfnamefont{F.}~\bibnamefont{Petruccione}},
  \emph{\bibinfo{title}{{The Theory of Open Quantum Systems}}}
  (\bibinfo{publisher}{Oxford University Press}, \bibinfo{year}{2002}).

\bibitem[{\citenamefont{Kossakowski}(1972)}]{Kossakowski1972}
\bibinfo{author}{\bibfnamefont{A.}~\bibnamefont{Kossakowski}},
  \bibinfo{journal}{Reports on Mathematical Physics}
  \textbf{\bibinfo{volume}{3}}, \bibinfo{pages}{247} (\bibinfo{year}{1972}).

\bibitem[{\citenamefont{Lindblad}(1976)}]{Lindblad1976}
\bibinfo{author}{\bibfnamefont{G.}~\bibnamefont{Lindblad}},
  \bibinfo{journal}{Communications in Mathematical Physics}
  \textbf{\bibinfo{volume}{48}}, \bibinfo{pages}{119} (\bibinfo{year}{1976}).

\bibitem[{\citenamefont{Judson and Rabitz}(1992)}]{Judson:92a}
\bibinfo{author}{\bibfnamefont{R.}~\bibnamefont{Judson}} \bibnamefont{and}
  \bibinfo{author}{\bibfnamefont{H.~A.} \bibnamefont{Rabitz}},
  \bibinfo{journal}{Physical Review Letters} \textbf{\bibinfo{volume}{68}},
  \bibinfo{pages}{1500} (\bibinfo{year}{1992}).

\bibitem[{\citenamefont{Rabitz et~al.}(2000)\citenamefont{Rabitz,
  de~Vivie-Riedle, Motzkus, and Kompa}}]{Rabitz2000}
\bibinfo{author}{\bibfnamefont{H.~A.} \bibnamefont{Rabitz}},
  \bibinfo{author}{\bibfnamefont{R.}~\bibnamefont{de~Vivie-Riedle}},
  \bibinfo{author}{\bibfnamefont{M.}~\bibnamefont{Motzkus}}, \bibnamefont{and}
  \bibinfo{author}{\bibfnamefont{K.}~\bibnamefont{Kompa}},
  \bibinfo{journal}{Science} \textbf{\bibinfo{volume}{288}},
  \bibinfo{pages}{824} (\bibinfo{year}{2000}).

\bibitem[{\citenamefont{Rabitz}(2003)}]{Rabitz:03a}
\bibinfo{author}{\bibfnamefont{H.~A.} \bibnamefont{Rabitz}},
  \bibinfo{journal}{Science} \textbf{\bibinfo{volume}{299}},
  \bibinfo{pages}{525} (\bibinfo{year}{2003}).

\bibitem[{\citenamefont{Zhu et~al.}(1998)\citenamefont{Zhu, Botina, and
  Rabitz}}]{Zhu1998}
\bibinfo{author}{\bibfnamefont{W.}~\bibnamefont{Zhu}},
  \bibinfo{author}{\bibfnamefont{J.}~\bibnamefont{Botina}}, \bibnamefont{and}
  \bibinfo{author}{\bibfnamefont{H.~A.} \bibnamefont{Rabitz}},
  \bibinfo{journal}{The Journal of Chemical Physics}
  \textbf{\bibinfo{volume}{108}}, \bibinfo{pages}{1953} (\bibinfo{year}{1998}).

\bibitem[{\citenamefont{Zhu and Rabitz}(1998)}]{Zhu1998a}
\bibinfo{author}{\bibfnamefont{W.}~\bibnamefont{Zhu}} \bibnamefont{and}
  \bibinfo{author}{\bibfnamefont{H.~A.} \bibnamefont{Rabitz}},
  \bibinfo{journal}{Journal of Chemical Physics}
  \textbf{\bibinfo{volume}{109}}, \bibinfo{pages}{385} (\bibinfo{year}{1998}).

\bibitem[{\citenamefont{Ohtsuki et~al.}(1999)\citenamefont{Ohtsuki, Zhu, and
  Rabitz}}]{Ohtsuki1999}
\bibinfo{author}{\bibfnamefont{Y.}~\bibnamefont{Ohtsuki}},
  \bibinfo{author}{\bibfnamefont{W.}~\bibnamefont{Zhu}}, \bibnamefont{and}
  \bibinfo{author}{\bibfnamefont{H.~A.} \bibnamefont{Rabitz}},
  \bibinfo{journal}{The Journal of Chemical Physics}
  \textbf{\bibinfo{volume}{110}}, \bibinfo{pages}{9825} (\bibinfo{year}{1999}).

\bibitem[{\citenamefont{Maday and Turinici}(2003)}]{maday2003}
\bibinfo{author}{\bibfnamefont{Y.}~\bibnamefont{Maday}} \bibnamefont{and}
  \bibinfo{author}{\bibfnamefont{G.}~\bibnamefont{Turinici}},
  \bibinfo{journal}{The Journal of Chemical Physics}
  \textbf{\bibinfo{volume}{118}}, \bibinfo{pages}{8191} (\bibinfo{year}{2003}).

\bibitem[{\citenamefont{Ohtsuki et~al.}(2004)\citenamefont{Ohtsuki, Turinici,
  and Rabitz}}]{Ohtsuki2004}
\bibinfo{author}{\bibfnamefont{Y.}~\bibnamefont{Ohtsuki}},
  \bibinfo{author}{\bibfnamefont{G.}~\bibnamefont{Turinici}}, \bibnamefont{and}
  \bibinfo{author}{\bibfnamefont{H.~A.} \bibnamefont{Rabitz}},
  \bibinfo{journal}{The Journal of Chemical Physics}
  \textbf{\bibinfo{volume}{120}}, \bibinfo{pages}{5509} (\bibinfo{year}{2004}).

\bibitem[{\citenamefont{Werschnik and Gross}(2007)}]{Werschnik2007}
\bibinfo{author}{\bibfnamefont{J.}~\bibnamefont{Werschnik}} \bibnamefont{and}
  \bibinfo{author}{\bibfnamefont{E.~K.~U.} \bibnamefont{Gross}},
  \bibinfo{journal}{Journal of Physics B: Atomic and Molecular Physics}
  \textbf{\bibinfo{volume}{40}}, \bibinfo{pages}{R175} (\bibinfo{year}{2007}).

\bibitem[{\citenamefont{Hartmann et~al.}(2014)\citenamefont{Hartmann, Latorre,
  Zhang, and Pavliotis}}]{Zhang2014}
\bibinfo{author}{\bibfnamefont{C.}~\bibnamefont{Hartmann}},
  \bibinfo{author}{\bibfnamefont{J.~C.} \bibnamefont{Latorre}},
  \bibinfo{author}{\bibfnamefont{W.}~\bibnamefont{Zhang}}, \bibnamefont{and}
  \bibinfo{author}{\bibfnamefont{G.~A.} \bibnamefont{Pavliotis}},
  \bibinfo{journal}{Journal of Computational Dynamics}
  \textbf{\bibinfo{volume}{1}}, \bibinfo{pages}{279} (\bibinfo{year}{2014}).

\bibitem[{\citenamefont{Benner et~al.}(2017)\citenamefont{Benner, Breiten,
  Hartmann, and Schmidt}}]{BSchmidt:77}
\bibinfo{author}{\bibfnamefont{P.}~\bibnamefont{Benner}},
  \bibinfo{author}{\bibfnamefont{T.}~\bibnamefont{Breiten}},
  \bibinfo{author}{\bibfnamefont{C.}~\bibnamefont{Hartmann}}, \bibnamefont{and}
  \bibinfo{author}{\bibfnamefont{B.}~\bibnamefont{Schmidt}},
  \bibinfo{journal}{SIAM Journal on Applied Dynamical Systems (SIADS),
  submitted. arXiv:1706.09882}  (\bibinfo{year}{2017}).

\bibitem[{\citenamefont{Sch{\"{a}}fer-Bung
  et~al.}(2011)\citenamefont{Sch{\"{a}}fer-Bung, Hartmann, Schmidt, and
  Sch{\"{u}}tte}}]{BSchmidt:61}
\bibinfo{author}{\bibfnamefont{B.}~\bibnamefont{Sch{\"{a}}fer-Bung}},
  \bibinfo{author}{\bibfnamefont{C.}~\bibnamefont{Hartmann}},
  \bibinfo{author}{\bibfnamefont{B.}~\bibnamefont{Schmidt}}, \bibnamefont{and}
  \bibinfo{author}{\bibfnamefont{C.}~\bibnamefont{Sch{\"{u}}tte}},
  \bibinfo{journal}{The Journal of Chemical Physics}
  \textbf{\bibinfo{volume}{135}}, \bibinfo{pages}{014112}
  (\bibinfo{year}{2011}).

\bibitem[{\citenamefont{Hartmann et~al.}(2010)\citenamefont{Hartmann, Vulcanov,
  and Sch{\"{u}}tte}}]{hartmann2010c}
\bibinfo{author}{\bibfnamefont{C.}~\bibnamefont{Hartmann}},
  \bibinfo{author}{\bibfnamefont{V.-M.} \bibnamefont{Vulcanov}},
  \bibnamefont{and}
  \bibinfo{author}{\bibfnamefont{C.}~\bibnamefont{Sch{\"{u}}tte}},
  \bibinfo{journal}{Multiscale Modeling {\&} Simulation}
  \textbf{\bibinfo{volume}{8}}, \bibinfo{pages}{1348} (\bibinfo{year}{2010}).

\bibitem[{\citenamefont{Hartmann et~al.}(2013)\citenamefont{Hartmann,
  Sch{\"{a}}fer-Bung, and Th{\"{o}}ns-Zueva}}]{hartmann2010e}
\bibinfo{author}{\bibfnamefont{C.}~\bibnamefont{Hartmann}},
  \bibinfo{author}{\bibfnamefont{B.}~\bibnamefont{Sch{\"{a}}fer-Bung}},
  \bibnamefont{and}
  \bibinfo{author}{\bibfnamefont{A.}~\bibnamefont{Th{\"{o}}ns-Zueva}},
  \bibinfo{journal}{SIAM Journal on Control and Optimization}
  \textbf{\bibinfo{volume}{51}}, \bibinfo{pages}{2356} (\bibinfo{year}{2013}).

\bibitem[{\citenamefont{Breiten and Damm}(2010)}]{Breiten2010}
\bibinfo{author}{\bibfnamefont{T.}~\bibnamefont{Breiten}} \bibnamefont{and}
  \bibinfo{author}{\bibfnamefont{T.}~\bibnamefont{Damm}},
  \bibinfo{journal}{Systems {\&} Control Letters}
  \textbf{\bibinfo{volume}{59}}, \bibinfo{pages}{443} (\bibinfo{year}{2010}).

\bibitem[{\citenamefont{Benner and Breiten}(2012)}]{Benner2012}
\bibinfo{author}{\bibfnamefont{P.}~\bibnamefont{Benner}} \bibnamefont{and}
  \bibinfo{author}{\bibfnamefont{T.}~\bibnamefont{Breiten}},
  \bibinfo{journal}{SIAM Journal on Matrix Analysis and Applications}
  \textbf{\bibinfo{volume}{33}}, \bibinfo{pages}{859} (\bibinfo{year}{2012}).

\bibitem[{\citenamefont{Korolkov et~al.}(1996)\citenamefont{Korolkov,
  Paramonov, and Schmidt}}]{BSchmidt:14}
\bibinfo{author}{\bibfnamefont{M.~V.} \bibnamefont{Korolkov}},
  \bibinfo{author}{\bibfnamefont{G.}~\bibnamefont{Paramonov}},
  \bibnamefont{and} \bibinfo{author}{\bibfnamefont{B.}~\bibnamefont{Schmidt}},
  \bibinfo{journal}{The Journal of Chemical Physics}
  \textbf{\bibinfo{volume}{105}}, \bibinfo{pages}{1862} (\bibinfo{year}{1996}).

\bibitem[{\citenamefont{Mecke}(1950)}]{Mecke:50a}
\bibinfo{author}{\bibfnamefont{R.}~\bibnamefont{Mecke}},
  \bibinfo{journal}{Zeitschrift f{\"{u}}r Elektrochemie und angewandte
  physikalische Chemie} \textbf{\bibinfo{volume}{54}}, \bibinfo{pages}{38}
  (\bibinfo{year}{1950}).

\bibitem[{\citenamefont{Owschimikow et~al.}(2011)\citenamefont{Owschimikow,
  Schmidt, and Schwentner}}]{BSchmidt:60}
\bibinfo{author}{\bibfnamefont{N.}~\bibnamefont{Owschimikow}},
  \bibinfo{author}{\bibfnamefont{B.}~\bibnamefont{Schmidt}}, \bibnamefont{and}
  \bibinfo{author}{\bibfnamefont{N.}~\bibnamefont{Schwentner}},
  \bibinfo{journal}{Physical Chemistry Chemical Physics}
  \textbf{\bibinfo{volume}{13}}, \bibinfo{pages}{8671} (\bibinfo{year}{2011}).

\bibitem[{\citenamefont{Schmidt and Friedrich}(2014)}]{BSchmidt:68}
\bibinfo{author}{\bibfnamefont{B.}~\bibnamefont{Schmidt}} \bibnamefont{and}
  \bibinfo{author}{\bibfnamefont{B.}~\bibnamefont{Friedrich}},
  \bibinfo{journal}{The Journal of Chemical Physics}
  \textbf{\bibinfo{volume}{140}}, \bibinfo{pages}{064317}
  (\bibinfo{year}{2014}).

\bibitem[{\citenamefont{Steinfeld et~al.}(1989)\citenamefont{Steinfeld,
  Francisco, and Hase}}]{Steinfeld:89a}
\bibinfo{author}{\bibfnamefont{J.~I.} \bibnamefont{Steinfeld}},
  \bibinfo{author}{\bibfnamefont{J.~S.} \bibnamefont{Francisco}},
  \bibnamefont{and} \bibinfo{author}{\bibfnamefont{W.~L.} \bibnamefont{Hase}},
  \emph{\bibinfo{title}{{Chemical Kinetics and Dynamics}}}
  (\bibinfo{publisher}{Prentice Hall}, \bibinfo{address}{Englewood Cliffs, NJ},
  \bibinfo{year}{1989}).

\bibitem[{\citenamefont{Burkey and Cantrell}(1984)}]{Burkey1984}
\bibinfo{author}{\bibfnamefont{R.~S.} \bibnamefont{Burkey}} \bibnamefont{and}
  \bibinfo{author}{\bibfnamefont{C.~D.} \bibnamefont{Cantrell}},
  \bibinfo{journal}{Journal of the Optical Society of America B}
  \textbf{\bibinfo{volume}{1}}, \bibinfo{pages}{169} (\bibinfo{year}{1984}).

\bibitem[{\citenamefont{Seel and Domcke}(1991)}]{Seel1991}
\bibinfo{author}{\bibfnamefont{M.}~\bibnamefont{Seel}} \bibnamefont{and}
  \bibinfo{author}{\bibfnamefont{W.}~\bibnamefont{Domcke}},
  \bibinfo{journal}{The Journal of Chemical Physics}
  \textbf{\bibinfo{volume}{95}}, \bibinfo{pages}{7806} (\bibinfo{year}{1991}).

\bibitem[{\citenamefont{Zhou and Doyle}(1998)}]{Zhou1998}
\bibinfo{author}{\bibfnamefont{K.}~\bibnamefont{Zhou}} \bibnamefont{and}
  \bibinfo{author}{\bibfnamefont{J.~C.} \bibnamefont{Doyle}},
  \emph{\bibinfo{title}{{Essentials of Robust Control}}}
  (\bibinfo{publisher}{Prentice Hall}, \bibinfo{year}{1998}).

\bibitem[{\citenamefont{Andrianov and Saalfrank}(2006)}]{Andrianov2006}
\bibinfo{author}{\bibfnamefont{I.}~\bibnamefont{Andrianov}} \bibnamefont{and}
  \bibinfo{author}{\bibfnamefont{P.}~\bibnamefont{Saalfrank}},
  \bibinfo{journal}{The Journal of Chemical Physics}
  \textbf{\bibinfo{volume}{124}}, \bibinfo{pages}{034710}
  (\bibinfo{year}{2006}).

\bibitem[{\citenamefont{Sundermann and de~Vivie-Riedle}(1999)}]{Sundermann:99a}
\bibinfo{author}{\bibfnamefont{K.}~\bibnamefont{Sundermann}} \bibnamefont{and}
  \bibinfo{author}{\bibfnamefont{R.}~\bibnamefont{de~Vivie-Riedle}},
  \bibinfo{journal}{The Journal of Chemical Physics}
  \textbf{\bibinfo{volume}{110}}, \bibinfo{pages}{1896} (\bibinfo{year}{1999}).

\bibitem[{\citenamefont{{Le Bris} et~al.}(2003)\citenamefont{{Le Bris}, Maday,
  and Turinici}}]{LeBris2003}
\bibinfo{author}{\bibfnamefont{C.}~\bibnamefont{{Le Bris}}},
  \bibinfo{author}{\bibfnamefont{Y.}~\bibnamefont{Maday}}, \bibnamefont{and}
  \bibinfo{author}{\bibfnamefont{G.}~\bibnamefont{Turinici}}, in
  \emph{\bibinfo{booktitle}{Quantum Control: Mathematical and Numerical
  Challenges : CRM Workshop}}, edited by \bibinfo{editor}{\bibfnamefont{A.~D.}
  \bibnamefont{Bandrauk}}, \bibinfo{editor}{\bibfnamefont{M.~C.}
  \bibnamefont{Delfour}}, \bibnamefont{and}
  \bibinfo{editor}{\bibfnamefont{C.}~\bibnamefont{{Le Bris}}}
  (\bibinfo{publisher}{American Chemical Society}, \bibinfo{year}{2003}), p.
  \bibinfo{pages}{139}.

\bibitem[{\citenamefont{Soml{\'{o}}i et~al.}(1993)\citenamefont{Soml{\'{o}}i,
  Kazakov, and Tannor}}]{somloi1993}
\bibinfo{author}{\bibfnamefont{J.}~\bibnamefont{Soml{\'{o}}i}},
  \bibinfo{author}{\bibfnamefont{V.~A.} \bibnamefont{Kazakov}},
  \bibnamefont{and} \bibinfo{author}{\bibfnamefont{D.~J.}
  \bibnamefont{Tannor}}, \bibinfo{journal}{Chemical Physics}
  \textbf{\bibinfo{volume}{172}}, \bibinfo{pages}{85} (\bibinfo{year}{1993}).

\bibitem[{\citenamefont{Khaneja et~al.}(2005)\citenamefont{Khaneja, Reiss,
  Kehlet, Schulte-Herbr{\"{u}}ggen, and Glaser}}]{Khaneja2005}
\bibinfo{author}{\bibfnamefont{N.}~\bibnamefont{Khaneja}},
  \bibinfo{author}{\bibfnamefont{T.}~\bibnamefont{Reiss}},
  \bibinfo{author}{\bibfnamefont{C.}~\bibnamefont{Kehlet}},
  \bibinfo{author}{\bibfnamefont{T.}~\bibnamefont{Schulte-Herbr{\"{u}}ggen}},
  \bibnamefont{and} \bibinfo{author}{\bibfnamefont{S.~J.}
  \bibnamefont{Glaser}}, \bibinfo{journal}{Journal of Magnetic Resonance}
  \textbf{\bibinfo{volume}{172}}, \bibinfo{pages}{296} (\bibinfo{year}{2005}).

\bibitem[{\citenamefont{Zhang and Lam}(2002)}]{Zhang:02a}
\bibinfo{author}{\bibfnamefont{L.}~\bibnamefont{Zhang}} \bibnamefont{and}
  \bibinfo{author}{\bibfnamefont{J.}~\bibnamefont{Lam}},
  \bibinfo{journal}{Automatica} \textbf{\bibinfo{volume}{38}},
  \bibinfo{pages}{205} (\bibinfo{year}{2002}).

\bibitem[{\citenamefont{Bai and Skoogh}(2006)}]{Bai2006}
\bibinfo{author}{\bibfnamefont{Z.}~\bibnamefont{Bai}} \bibnamefont{and}
  \bibinfo{author}{\bibfnamefont{D.}~\bibnamefont{Skoogh}},
  \bibinfo{journal}{Linear Algebra and its Applications}
  \textbf{\bibinfo{volume}{415}}, \bibinfo{pages}{406} (\bibinfo{year}{2006}).

\bibitem[{\citenamefont{Wachspress}(1988)}]{Wachspress1988}
\bibinfo{author}{\bibfnamefont{E.~L.} \bibnamefont{Wachspress}},
  \bibinfo{journal}{Applied Mathematics Letters} \textbf{\bibinfo{volume}{1}},
  \bibinfo{pages}{87} (\bibinfo{year}{1988}).

\bibitem[{\citenamefont{Damm}(2008)}]{Damm2008}
\bibinfo{author}{\bibfnamefont{T.}~\bibnamefont{Damm}},
  \bibinfo{journal}{Numerical Linear Algebra with Applications}
  \textbf{\bibinfo{volume}{15}}, \bibinfo{pages}{853} (\bibinfo{year}{2008}).

\bibitem[{\citenamefont{Laub et~al.}(1987)\citenamefont{Laub, Heath, Paige, and
  Ward}}]{Laub1987}
\bibinfo{author}{\bibfnamefont{A.}~\bibnamefont{Laub}},
  \bibinfo{author}{\bibfnamefont{M.}~\bibnamefont{Heath}},
  \bibinfo{author}{\bibfnamefont{C.}~\bibnamefont{Paige}}, \bibnamefont{and}
  \bibinfo{author}{\bibfnamefont{R.}~\bibnamefont{Ward}},
  \bibinfo{journal}{IEEE Transactions on Automatic Control}
  \textbf{\bibinfo{volume}{32}}, \bibinfo{pages}{115} (\bibinfo{year}{1987}).

\bibitem[{\citenamefont{Tombs and Postlethwaite}(1987)}]{Tombs1987}
\bibinfo{author}{\bibfnamefont{M.~S.} \bibnamefont{Tombs}} \bibnamefont{and}
  \bibinfo{author}{\bibfnamefont{I.}~\bibnamefont{Postlethwaite}},
  \bibinfo{journal}{International Journal of Control}
  \textbf{\bibinfo{volume}{46}}, \bibinfo{pages}{1319} (\bibinfo{year}{1987}).

\bibitem[{\citenamefont{Hahn and Edgar}(2002)}]{Hahn2002}
\bibinfo{author}{\bibfnamefont{J.}~\bibnamefont{Hahn}} \bibnamefont{and}
  \bibinfo{author}{\bibfnamefont{T.~F.} \bibnamefont{Edgar}},
  \bibinfo{journal}{Industrial {\&} Engineering Chemistry Research}
  \textbf{\bibinfo{volume}{41}}, \bibinfo{pages}{2204} (\bibinfo{year}{2002}).

\bibitem[{\citenamefont{Horenko
  et~al.}(2002{\natexlab{a}})\citenamefont{Horenko, Schmidt, and
  Sch{\"{u}}tte}}]{BSchmidt:38}
\bibinfo{author}{\bibfnamefont{I.}~\bibnamefont{Horenko}},
  \bibinfo{author}{\bibfnamefont{B.}~\bibnamefont{Schmidt}}, \bibnamefont{and}
  \bibinfo{author}{\bibfnamefont{C.}~\bibnamefont{Sch{\"{u}}tte}},
  \bibinfo{journal}{The Journal of Chemical Physics}
  \textbf{\bibinfo{volume}{117}}, \bibinfo{pages}{4643}
  (\bibinfo{year}{2002}{\natexlab{a}}).

\bibitem[{\citenamefont{Horenko
  et~al.}(2002{\natexlab{b}})\citenamefont{Horenko, Salzmann, Schmidt, and
  Sch{\"{u}}tte}}]{BSchmidt:39}
\bibinfo{author}{\bibfnamefont{I.}~\bibnamefont{Horenko}},
  \bibinfo{author}{\bibfnamefont{C.}~\bibnamefont{Salzmann}},
  \bibinfo{author}{\bibfnamefont{B.}~\bibnamefont{Schmidt}}, \bibnamefont{and}
  \bibinfo{author}{\bibfnamefont{C.}~\bibnamefont{Sch{\"{u}}tte}},
  \bibinfo{journal}{The Journal of Chemical Physics}
  \textbf{\bibinfo{volume}{117}}, \bibinfo{pages}{11075}
  (\bibinfo{year}{2002}{\natexlab{b}}).

\bibitem[{\citenamefont{Horenko et~al.}(2004)\citenamefont{Horenko, Weiser,
  Schmidt, and Sch{\"{u}}tte}}]{BSchmidt:43}
\bibinfo{author}{\bibfnamefont{I.}~\bibnamefont{Horenko}},
  \bibinfo{author}{\bibfnamefont{M.}~\bibnamefont{Weiser}},
  \bibinfo{author}{\bibfnamefont{B.}~\bibnamefont{Schmidt}}, \bibnamefont{and}
  \bibinfo{author}{\bibfnamefont{C.}~\bibnamefont{Sch{\"{u}}tte}},
  \bibinfo{journal}{The Journal of Chemical Physics}
  \textbf{\bibinfo{volume}{120}}, \bibinfo{pages}{8913} (\bibinfo{year}{2004}).

\bibitem[{\citenamefont{Subotnik et~al.}(2016)\citenamefont{Subotnik, Jain,
  Landry, Petit, Ouyang, and Bellonzi}}]{Subotnik2016}
\bibinfo{author}{\bibfnamefont{J.~E.} \bibnamefont{Subotnik}},
  \bibinfo{author}{\bibfnamefont{A.}~\bibnamefont{Jain}},
  \bibinfo{author}{\bibfnamefont{B.}~\bibnamefont{Landry}},
  \bibinfo{author}{\bibfnamefont{A.}~\bibnamefont{Petit}},
  \bibinfo{author}{\bibfnamefont{W.}~\bibnamefont{Ouyang}}, \bibnamefont{and}
  \bibinfo{author}{\bibfnamefont{N.}~\bibnamefont{Bellonzi}},
  \bibinfo{journal}{Annual Review of Physical Chemistry}
  \textbf{\bibinfo{volume}{67}}, \bibinfo{pages}{387} (\bibinfo{year}{2016}).

\bibitem[{\citenamefont{Tremblay et~al.}(2011)\citenamefont{Tremblay,
  Klinkusch, Klamroth, and Saalfrank}}]{Tremblay2011a}
\bibinfo{author}{\bibfnamefont{J.~C.} \bibnamefont{Tremblay}},
  \bibinfo{author}{\bibfnamefont{S.}~\bibnamefont{Klinkusch}},
  \bibinfo{author}{\bibfnamefont{T.}~\bibnamefont{Klamroth}}, \bibnamefont{and}
  \bibinfo{author}{\bibfnamefont{P.}~\bibnamefont{Saalfrank}},
  \bibinfo{journal}{Journal of Chemical Physics}
  \textbf{\bibinfo{volume}{134}}, \bibinfo{pages}{044311}
  (\bibinfo{year}{2011}).

\bibitem[{\citenamefont{Lockwood et~al.}(2001)\citenamefont{Lockwood, Ratner,
  and Kosloff}}]{Lockwood2001}
\bibinfo{author}{\bibfnamefont{D.~M.} \bibnamefont{Lockwood}},
  \bibinfo{author}{\bibfnamefont{M.}~\bibnamefont{Ratner}}, \bibnamefont{and}
  \bibinfo{author}{\bibfnamefont{R.}~\bibnamefont{Kosloff}},
  \bibinfo{journal}{Chemical Physics} \textbf{\bibinfo{volume}{268}},
  \bibinfo{pages}{55} (\bibinfo{year}{2001}).

\end{thebibliography}

\clearpage

\clearpage

\clearpage

\clearpage

\clearpage

\end{document}